# Contradictions

Yang Xu[a, c*], Shuwei Chen[a, c*], Xiaomei Zhong[a, c], Jun Liu[b, c], and Xingxing He[a, c]

[a] *School of Mathematics, Southwest Jiaotong University, Chengdu, Sichuan, China*

[b] *School of Computing, Faculty of Computing, Engineering and the Built Environment, Ulster University, Belfast, Northern Ireland, UK*

[c] *National-Local Joint Engineering Laboratory of System Credibility Automatic Verification, Southwest Jiaotong University, Chengdu, Sichuan, China*

**Abstract:** Trustworthy AI requires reasoning systems that are not only powerful but also transparent and reliable. Automated Theorem Proving (ATP) is central to formal reasoning, yet classical binary resolution remains limited, as each step involves only two clauses and eliminates at most two literals. To overcome this bottleneck, the concept of *standard contradiction* and the theory of contradiction-separation-based deduction were introduced in 2018. This paper advances that framework by focusing on the systematic construction of standard contradictions. Specially, this study investigates construction methods for two principal forms of standard contradiction: the maximum triangular standard contradiction and the triangular-type standard contradiction. Building on these structures, we propose a procedure for determining the satisfiability and unsatisfiability of clause sets via maximum standard contradiction. Furthermore, we derive formulas for computing the number of standard sub-contradictions embedded within both the maximum triangular standard contradiction and the triangular-type standard contradiction. The results presented herein furnish the methodological basis for advancing contradiction-separation-based dynamic multi-clause automated deduction, thereby extending the expressive and deductive capabilities of automated reasoning systems beyond the classical binary paradigm.

## 1. Introduction

Trustworthy Artificial Intelligence (AI) requires reasoning systems that are not only powerful but also reliable, transparent, and adaptable. Among the core pillars of trustworthy AI are automatic reasoning and knowledge representation, which together provide interpretable and verifiable mechanisms for intelligent decision-making. Recent initiatives emphasize that AI systems must demonstrate robustness, transparency, and accountability if they are to be trusted in high-stakes domains. Logic plays a central role in this context, offering a robust and

* Corresponding authors.

All the five authors are co-first authors.

*E-mail addresses*: xuyang@swjtu.edu.cn (Y. Xu), swchen@swjtu.edu.cn (S. Chen), zhongxm2013@swjtu.edu.cn (X. Zhong), j.liu@ulster.ac.uk (J. Liu), x.he@swjtu.edu.cn (X. He)

expressive language for structuring knowledge in ways that are both human-understandable and machine-processable. Automated Theorem Proving (ATP) is one of the most prominent tools in this domain. Based on first-order logic, ATP systems use symbolic reasoning to derive conclusions with mathematical rigor, making them essential for applications such as mathematical proof verification, problem solving, program verification, and broader domains of formal verification and safety-critical AI.

Since Hao Wang first derived inference rules from logical deduction, used computers to prove 220 theorems from *Principia Mathematica* in 1959, and subsequently published *Toward Mechanical Mathematics* [1] in the *IBM Journal of Research and Development*, logic-based automated reasoning systems have developed rapidly. A major breakthrough came in 1965, when Robinson proposed the Resolution Principle [2]. ATP systems built upon this principle quickly became one of the most influential and widely applied directions in automated deductive reasoning, producing a wealth of important results. This breakthrough method for first-order logic forms the theoretical basis for modern ATP systems. State-of-the-art systems—including Vampire [3,19], E [20], SPASS [21], and Prover9 [22]—still rely on binary resolution as their core inference mechanism [23–25].

Nevertheless, the capability and efficiency of resolution-based ATP systems remain insufficient for many practical demands. For example, among the most challenging conjunctive normal form (CNF) problems (Rating 1) in the well-known TPTP problem library [3], 1,581 instances can be solved only by a very small number of ATP systems. The underlying limitation lies in the fundamental mechanism of **binary resolution**, wherein each inference step involves exactly two clauses and eliminates at most two literals. Although variants such as group resolution [4], linear resolution [5, 6], and lock resolution [7] have been proposed, they still conform to this binary restriction, thereby constraining further improvements in both capability and efficiency.

To overcome this bottleneck, Xu *et al.* [8] introduced the concept of **contradiction separation (CS)** and proposed a novel inference rule for automated deduction. This approach generalizes classical binary resolution to handle multiple clauses dynamically and synergistically, with binary resolution appearing as a special case. CS-based automated deduction is thus characterized as a **dynamic, multi-clause, collaborative, goal-oriented, and robust framework** [9]. This advancement provides a more powerful and efficient methodology for addressing logical reasoning and problem-solving challenges across scientific research, technological innovation, and social applications.

Based on this framework, corresponding methods and systems for CS-based dynamic automated deduction have been developed [10–13]. These systems have collectively proved over 260,000 theorems expressed in first-order logic. Among them, the prover CSI-V1.1, which integrates CS-based inference with Vampire 4.9—the most advanced ATP system of the past two decades—has proved **11498** theorems from TPTP, exceeding Vampire 4.9 alone by **470**. A comparative evaluation against Vampire 4.9 has also been conducted using problems from the CASC2020 to CASC2024 (500 problems per year), with results summarized in Table 1.1. Moreover, by combining our prover with the internationally renowned prover E, we have achieved outstanding performance in CASC since 2018, including one second-place finish, three third places, two fourth places, and one fifth place in the FOF division, as well as a first-ever second place in the ICU division. Additionally, we were awarded the "Newcomer Prize" in 2018.

Table 1.1. Comparison results of CSI-V 1.1 against Vampire 4.9

| | 2020 (500) | 2021 (500) | 2022 (500) | 2023 (500) | 2024 (500) | Total (2500) |
|---|---|---|---|---|---|---|
| Vampire 4.9 | 457 | 455 | 453 | 450 | 451 | 2266 |
| CSI-V 1.1 | **487** | **485** | **492** | **491** | **491** | **2446** |

In contradiction-separation-based dynamic deduction, the construction of contradictions constitutes the essential mechanism for generating new separation clauses. Unlike classical binary resolution, where each inference step is strictly limited to two clauses and at most two literals, this framework permits the participating clauses and literals in a contradiction to be flexibly and dynamically varied during the deduction process. Such flexibility greatly enhances the expressive and deductive capacity of the system but also makes the method of contradiction construction a decisive factor in determining overall efficiency. This paper focuses primarily on the core theory and methods of contradiction-separation-based dynamic automated deduction, with particular emphasis on the **theory and methods for constructing and using contradictions determining satisfiability and unsatisfiability of clause sets**. In particular, different construction strategies yield different deductive behaviors. We introduce the *maximal contradiction* provides a comprehensive, static structure capable of determining satisfiability and unsatisfiability of clause sets, whereas the *triangular standard contradiction* introduces a dynamic mechanism that allows contradictions to be constructed and adjusted simultaneously with the deduction process itself. Thus, the study of contradiction construction methods—both maximal and triangular—forms the theoretical and methodological foundation for advancing contradiction-separation-based dynamic multi-clause deduction.

This paper makes the following key contributions:

1) **Definition, Construction and Use of the Maximal Contradiction**: We define the maximal contradiction, present its construction method, and provide theoretical proofs of its soundness. We establish its role in determining satisfiability and unsatisfiability of clause sets, and further show that it can generate concrete satisfiable instances when clause sets are satisfiable.

2) **Development of Triangular Standard Contradictions**: To overcome the rigidity of maximal contradictions, we introduce triangular standard contradictions, which allow contradictions to be dynamically adjusted during deduction. We present their formal definition, construction method, and theoretical validation.

3) **Analysis of Sub-Contradictions**: We study the structural properties of triangular standard contradictions by classifying their sub-contradictions into homotypic and non-homotypic types. Four forms of homotypic sub-contradictions are identified and analyzed, and formulas are derived for computing the number of sub-contradictions in both maximal and triangular standard contradictions.

4) **Methodological Foundation for Multi-Clause Deduction**: Collectively, the concepts, construction methods, and structural analyses introduced in this paper establish the methodological groundwork for advancing contradiction-separation-based dynamic multi-clause automated deduction, enabling more expressive and efficient reasoning systems.

The remainder of this paper is organized as follows: Section 2 provides an overview of related works. Section 3 introduces basic definitions and theoretical concepts of contradiction, forming the foundation for multi-clause automated deduction; Section 4 defines the maximal contradiction, presents its construction method, and establishes theoretical results on its relationship with satisfiability and unsatisfiability of clause sets. Methods for satisfiability determination and instance construction are described and illustrated with examples. Section 5 proposes triangular standard contradictions as a dynamic alternative to maximal contradictions. The definition, construction methods, and theoretical proofs of validity are provided, along with illustrative examples. In addition, Section 5 also examines the internal structure of triangular standard contradictions, classifies their sub-contradictions, and derives formulas for computing the number of sub-contradictions in both maximal and triangular standard contradictions. Section 6 summarizes the main contributions and outlines directions for future research, particularly the progressive development of contradiction-separation-based dynamic multi-clause deduction methods and their applications in automated theorem proving.

## 2. Related Works

### 2.1 Theories of Automated Deduction Based on the Resolution Principle

In 1965, Robinson proposed the resolution principle and unification [2], thereby laying the foundation for logic-based resolution automated deduction. To this day, the resolution principle remains the most influential and widely applied inference rule, dominating logic-based automated reasoning for decades.

In 1967, Slagle introduced group resolution through his theory of semantic clash deduction [4]. However, each "clash" in this approach is still realized through a sequence of binary resolution steps.

In 1968, Robinson [14] and later Harrison *et al.* (1978) [15] independently proposed the generalized resolution principle. Despite some theoretical novelty, these approaches shared two strict requirements: (1) all participating clauses must be binary, and (2) there must exist a special clause containing the negation of one literal from these binary clauses. Under these conditions, both principles produce a resolvent of essentially the same form. Although this enabled, in theory, the simultaneous handling of multiple clauses, in practice such special clauses were difficult to obtain. Moreover, the reliability of the method still relied fundamentally on binary resolution. Consequently, generalized resolution did not exert lasting influence.

Further attempts at extending resolution include Krishnamurthy (1985) [16], who combined the extension principle, the symmetry principle, and resolution to improve deductive capability, and Egly *et al.* (1993) [17], who developed *SR-deduction* by incorporating symmetry-derived clauses. In 2004, Tammet [18] proposed the chain resolution principle, where binary clauses are connected via implication chains; however, this principle effectively reduces to repeated applications of binary resolution.

In summary, since Robinson's seminal proposal, there has been no fundamental theoretical breakthrough in logic-based automated deduction for more than half a century. State-of-the-art systems—including Vampire [3, 19], E [20], SPASS [21], and Prover9 [22]—still rely on binary resolution as their core inference mechanism [23–25]. As early as 2003, Andrei Voronkov, one of the leading figures in automated reasoning and the developer of Vampire (a multiple-time CASC champion), emphasized the urgent need for new theoretical advances in logic-based deduction [26]. Yet, to date, no milestone theory has emerged [27].

### 2.2 Methods of Automated Deduction Based on the Resolution Principle

Although no breakthrough theories have been achieved since Robinson, a variety of **methods** have been proposed to enhance the efficiency of resolution-based deduction by optimizing the use of binary resolution.

**(a) Methods with Restricted Clauses or Literals**

Efforts in this direction focus on constraining clause interactions to reduce redundancy. Examples include:

- The set-of-support method (Wos *et al.*, 1965) [28], which restricts resolution by forbidding inference between clauses within a designated satisfiable subset.
- Semantic clash resolution (Slagle, 1967) [4], which partitions the clause set into disjoint subsets using an interpretation, prohibiting intra-subset resolution.
- Linear resolution (Luckham and Loveland, 1968) [5,6], which uses the derived resolvent as the pivot for subsequent resolutions; complete for propositional logic but not for first-order logic.
- Lock resolution (Boyer, 1971) [7], sound and complete but lacking a practical method for assigning locks.
- SLD resolution (Kowalski and Kuehner, 1971) [29], a goal-directed variant of linear resolution, complete under a selection function.
- Ordered linear resolution (Chang and Lee, 1973) [30], which improves efficiency but sacrifices completeness.
- Resolution with constrained clauses (Hans *et al.*, 1994) [31], sound and complete under specific restrictions.
- Goal-oriented semantic forward reasoning (Choi, 2002) [32], which minimizes redundant models before applying resolution.
- Generalized ordered resolution (Belaid *et al.*, 2010) [33].
- Goal-sensitive semantic resolution (Bonacina *et al.*, 2017) [34].
- Conflict resolution (Bruno *et al.*, 2017) [35,36], inspired by CDCL, which incorporates decision literals and conflict clauses but remains limited for first-order logic.

These approaches reduce redundancy and improve efficiency by selectively restricting clause and literal participation.

**(b) Methods Applying Heuristic Strategies**

Heuristics have also been widely used to guide resolution. For example, Fellows *et al.* (2006) [37] studied parameterized problem control to obtain bounded-length resolutions and minimally unsatisfiable subsets. Heuristic strategies have since been refined in major theorem provers such as SPASS, Vampire, Otter, and E.

**(c) Methods for Reducing the Deductive Search Space**

Another key direction has been to reduce the search space of resolution. Examples include:

- The layered resolution method (Anatoli *et al.*, 2003) [38], which eliminates redundant clauses but is limited in generality.
- Relevance filtering (Meng *et al.*, 2009) [36], which retains only clauses relevant to a strategically chosen subset.
- Splitting resolution (Prcovice, 2012) [39], evolved from extended resolution, which substantially reduces the combinatorial search space.
- Heuristic handling of inequalities (Avigad *et al.*, 2016) [40], demonstrating tailored heuristics for specific clause sets.

In summary, across these decades of work, numerous strategies have been devised to restrict clause interactions, prioritize specific inferences, apply heuristics, or reduce the search space. While these approaches have yielded notable improvements in efficiency, they all remain confined to the framework of **binary resolution**. To date, no method has succeeded in fundamentally breaking through the limitations imposed by the binary resolution principle.

In contrast, Xu *et al.* [8] advanced a new line of research through **contradiction separation (CS)** and the associated concept of **standard contradictions**, which enable the simultaneous participation of multiple clauses and the elimination of multiple literals in a single inference step. By shifting from binary to dynamic multi-clause deduction, this work addresses the limitations identified in prior research and establishes a new theoretical foundation for the next generation of logic-based automated reasoning.

## 3. Concepts of Contradiction

To overcome the limitations of binary resolution, deduction must allow multiple clauses to participate and multiple literals to be eliminated in a single step. This section introduces the fundamental concepts of *contradiction*, which provide the theoretical basis for such multi-clause deduction.

In propositional logic, *a literal* refers to a propositional variable or its negation. For example, let $l$ be a propositional variable in propositional logic; then both $l$ and $\sim l$ (the negation of $l$) are literals, and $l$ and $\sim l$ are called *a complementary pair*. A disjunction of several literals is called *a clause*, and a conjunction of several clauses is called a *conjunctive normal form* (CNF). For example, $S = C_1 \wedge C_2 \wedge \ldots \wedge C_m$ is a conjunctive normal form, where $C_1, C_2, \ldots, C_m$ are clauses. Sometimes $S$ is denoted as $S = \{C_1, C_2, \ldots, C_m\}$, in which case $S$ is also referred to as *a clause set*.

In first-order logic, if $P(x_1, \ldots, x_n)$ is an $n$-ary predicate symbol and $t_1, \ldots, t_n$ are terms, then

$P(t_1,\ldots, t_n)$ is *an atom*. A literal refers to an atom or its negation. A clause is a disjunction of several literals. A clause set $S$ is a predicate logic formula in which all clauses in $S$ are conjoined, and every variable in $S$ is considered to be universally quantified. For a clause $C$ in propositional logic or first-order logic, the set of all literals in $C$ is still denoted as $C$.

**Definition 3.1** [8] **(Contradiction)** Assume a clause set $S=\{C_1, C_2, \ldots, C_m\}$ in propositional logic or first-order logic. If for any literal $x_i$ from a clause $C_i$ ($i$=1,…, $m$), there exist some complementary pair of literals among $x_1,\ldots, x_m$, then $S=\bigwedge_{i=1}^{m} C_i$ is called a *standard contradiction*. If $\bigwedge_{i=1}^{m} C_i$ is unsatisfiable, then $S=\bigwedge_{i=1}^{m} C_i$ is called a *quasi-contradiction*, where $C_i$ is also regarded as a set of literals ($i$=1, …, $m$).

**Example 3.1** Suppose that $S=\{C_1, C_2, C_3, C_4\}$ is a clause set in propositional logic, where $C_1 = l_1 \vee l_2$, $C_2 = \sim l_2$, $C_3 = \sim l_1$, with $l_1$, $l_2$ being propositional variables. It can be seen that S is a standard contradiction, as well as a quasi-contradiction.

**Lemma 3.1** [8] Assume a clause set $S=\{C_1, C_2, \ldots, C_m\}$ in propositional logic, then $S$ is a standard contradiction if and only if $S$ is a quasi-contradiction. Therefore, in propositional logic, both standard contradiction and quasi-contradiction are shortly called *contradiction*.

**Remark 3.1** (1) Whether $\bigwedge_{i=1}^{m} C_i$ is a standard contradiction is regardless of the ordering of the clauses $C_1, C_2,\ldots, C_m$.

(2) In first-order logic, a quasi-contradiction is not necessarily a standard contradiction. Take the clause set $S=\{P(x), \sim P(f(y))\}$ as an example. It can be seen easily that $S$ is unsatisfiable, i.e., $S$ is a quasi-contradiction. However, $S$ is not a standard contradiction since $P(x)$ and $\sim P(f(y))$ are not complementary to each other.

**Lemma 3.2** [8] Assume a clause set $S=\{C_1, C_2, \ldots, C_m\}$ in first-order logic, if $S$ is a standard contradiction, then $S$ is a quasi-contradiction, i.e., $S$ is unsatisfiable.

**Corollary 3.1** [8] Suppose $S=\{C_1, C_2,\ldots, C_m\}$, where $C_1, C_2,\ldots, C_m$ are clauses in the first-order logic. If $C_1\wedge C_2\wedge\ldots\wedge C_m$ is a standard contradiction, then for any variable substitution $\sigma$ of $S$, $(C_1\wedge C_2\wedge\ldots\wedge C_m)^\sigma$ is still a standard contradiction.

**Remark 3.2** In first-order logic, a quasi-contradiction may not be a quasi-contradiction any more after applying some substitution. For example, considering $S=\{L(x), \sim L(f(a))\}$, here $a$ is a constant, obviously $S$ is unsatisfiable. Suppose a substitution $\sigma=\{a/x\}$ is applied, then $S^\sigma = \{L(a), \sim L(f(a))\}$. Obviously, $S^\sigma$ is satisfiable, and not a quasi-contradiction anymore.

**Theorem 3.1** Assume a standard contradiction $S=\{C_1, C_2, \ldots, C_m\}$ in propositional logic or first-order logic, if $C_i^*$ is a clause formed by some of the literals from $C_i$ ($i$=1, …, $m$), then $\bigwedge_{i=1}^{m} C_i^*$ is a standard contradiction.

*Proof.* Because $S=\{C_1, C_2, \ldots, C_m\}$ is a standard contradiction, for any literal $x_i$ in $C_i$, $i$ =

1, 2,…, $m$, there must be complementary pair of literals among $x_1,\ldots, x_m$. Furthermore, as $C_i^*$ is a clause formed by some of the literals from $C_i$ ($i$=1, …, $m$), then for any literal $y_i$ in $C_i^*$, there must be complementary pair of literals among $y_1,\ldots, y_m$, and therefore, $\wedge_{i=1}^{m} C_i^*$ is a standard contradiction. ■

In order to study the sub-structure of standard contradiction, the concept of standard sub-contradiction of a standard contradiction is given in the following Definition 3.2.

**Definition 3.2 (Sub-contradiction)** Assume $S$=$\wedge_{i=1}^{m} C_i$ is a standard contradiction in propositional logic or first-order logic. For $u$ = 1, 2,…, $t$, $C_u^*(i)$ is a clause formed by some of the literals from $C_i$, where $i \in \{1, 2,\ldots, m\}$, $t \le m$. If $\wedge_{u=1}^{t} C_u^*(i)$ is a standard contradiction, then $\wedge_{u=1}^{t} C_u^*(i)$ is called *a standard sub-contradiction* of $\wedge_{i=1}^{m} C_i$.

**Example 3.2** Suppose that $S$={$C_1$, $C_2$, $C_3$, $C_4$} is a standard contradiction in propositional logic, where $C_1 = l_1 \vee l_2$, $C_2 = l_1 \vee \sim l_2$, $C_3 = \sim l_1 \vee l_2$, $C_4 = \sim l_1 \vee \sim l_2$, with $l_1$, $l_2$ being propositional variables. Suppose that $C_1^* = l_1 \vee l_2$, $C_2^* = \sim l_2$, $C_3^* = \sim l_1$, then $C_1^* \wedge C_2^* \wedge C_3^*$ is a standard sub-contradiction of $S$.

**Definition 3.3** [8] Suppose a clause set $S = \{C_1, C_2, \ldots, C_m\}$ in first-order logic. Without loss of generality, assume that there does not exist the same variables among $C_1, C_2, \ldots, C_m$(if there exist the same variables, there exists a rename substitution can then be applied to make them different). The following inference rule that produces a new clause from $S$ is called a *standard contradiction separation rule*, in short, an S-CS rule:

For each $C_i$ ($i = 1,2,\ldots,m$), firstly apply a substitution $\sigma_i$ to $C_i$ ($\sigma_i$ could be an empty substitution but not necessary the most general unifier), denoted as $C_i^{\sigma_i}$; then separate $C_i^{\sigma_i}$ into two sub-clauses $C_i^{\sigma_i -}$ and $C_i^{\sigma_i +}$ such that

(1) $C_i^{\sigma_i} = C_i^{\sigma_i -} \vee C_i^{\sigma_i +}$, where $C_i^{\sigma_i -}$ and $C_i^{\sigma_i +}$ have no common literal;

(2) $C_i^{\sigma_i +}$ can be an empty clause itself, but $C_i^{\sigma_i -}$ cannot be an empty clause;

(3) $\wedge_{i=1}^{m} C_i^{\sigma_i -}$ is a standard contradiction, that is $\forall (x_1, \ldots, x_m) \in \prod_{i=1}^{m} C_i^{\sigma_i -}$, there exists at least one complementary pair among $\{x_1, \ldots, x_m\}$.

The resulting clause $\vee_{i=1}^{m} C_i^{\sigma_i +}$, denoted as $C_m^{s\sigma}(C_1, \ldots, C_m)$ (here "s" means "standard", $\sigma = \cup_{i=1}^{m} \sigma_i$, $\sigma_i$ is a substitution to $C_i$, $i = 1, \ldots, m$), is called a *standard contradiction separation clause* (S-CSC) of $C_1, \ldots, C_m$, and $\wedge_{i=1}^{m} C_i^{\sigma_i -}$ is called a *separated standard contradiction* (S-SC).

The most intuitive feature of this definition is the involvement of multiple clauses in the deduction, as reflected in $C_m^{s\sigma}(C_1, \ldots, C_m)$, reflecting the synergy of multiple clauses. We can see that the contradiction contains at least one complementary pair, as demonstrated in (3) of Definition 3.3. Typically, more than one complementary pair is deleted in the deduction process.

This is different to binary resolution.

**Definition 3.4** [8] Suppose a clause set $S = \{C_1, C_2, \ldots, C_m\}$ in first-order logic. $\Phi_1$,$\Phi_2$,...,$\Phi_t$ is called *a standard contradiction separation based dynamic deduction sequence from S to a clause* $\Phi_t$, denoted as $D^s$, if

(1) $\Phi_i \in S,\ i \in \{1,2, \ldots ,t\}$; or

(2) there exist $r_1,\ r_2,\ \ldots,\ r_{k_i} < i,\ \Phi_i = C^{s\theta_i}_{r_{k_i}}(\Phi_{r_1}, \Phi_{r_2}, \ldots , \Phi_{r_{k_i}})$, where $\theta_i = \bigcup_{j=1}^{k_i} \sigma_j$, $\sigma_j$ is a substitution to $\Phi_{r_j}$, $j = 1,2,\ldots ,k_i$.

In Definition 3.4, subscripts and substitutions can be changed, reflecting the dynamic and flexible feature of S-CS deduction. Of course, the S-CS rule also satisfies soundness and completeness, and the corresponding conclusions are as follows.

**Theorem 3.2** [8] (**Soundness Theorem of the S-CS Based Dynamic Deduction in First-Order Logic**) Suppose a clause set $S = \{C_1, C_2, \ldots, C_m\}$ in first-order logic. $\Phi_1$,$\Phi_2$,$\cdots$,$\Phi_t$ is an S-CS based dynamic deduction from S to a clause $\Phi_t$. If $\Phi_t$ is an empty clause, then $S$ is unsatisfiable.

**Theorem 3.3 [8]** (**Completeness of the S-CS Based Dynamic Deduction in First-Order Logic**) Suppose a clause set $S = \{C_1, C_2, \ldots, C_m\}$ in first-order logic. If $S$ is unsatisfiable, then there exists an S-CS based dynamic deduction from $S$ to an empty clause.

In summary, the **S-CS** rule offers several key advantages over traditional binary resolution:

(1) **Multi-Clause Involvement**: Unlike binary resolution, S-CS allows multiple clauses to be used simultaneously in a deduction. This means it isn't limited to using only two clauses that contain a complementary unifiable literal pair.

(2) **Dynamic, Synergized, and Flexible**: The S-CS rule's design makes the deduction process more adaptable and cooperative, allowing clauses to work together in a more integrated manner.

(3) **Controllability**: S-CS allows for direct control over the number of participating clauses and the size of the standard contradiction. This controllability helps to reduce the number of literals in the resulting clause, which effectively mitigates the space explosion problem that often challenges binary resolution-based systems.

(4) **Repeatability**: Clauses can be reused in the deduction process to speed up the overall proof search.

The work in [8] lays the first formal foundation for contradiction-separation-based dynamic multi-clause deduction, extending the theoretical scope of logic-based automated reasoning beyond binary resolution. By enabling simultaneous multi-clause participation, contradiction separation and standard contradictions open a new path toward significantly more efficient automated deduction. However, [8] is mainly placed a theoretical foundation, theory and methods for constructing and using contradictions are crucial, which is the focus of the present work, and are detailed in the subsequent sections.

## 4. Definition, Construction and Use of the Maximal Contradiction

To investigate methods for determining the properties of a clause set using contradictions, this section examines how the properties of a given clause set SSS can be established through a special contradiction constructed from all propositional variables occurring in SSS, namely the *maximal contradiction*.

The following presents the definition and construction method of the maximal contradiction, along with its theoretical proof.

**Theorem 4.1** Let $V=\{l_1,\ldots, l_n\}$ be a set of propositional variables, and for each variable $l_i$, let $p(i)\in\{l_i, \sim l_i\}$ be a literal. Suppose we have a clause $C(p(1),\ldots, p(i),\ldots, p(n))=\vee_{i=1}^{n}p(i)$, and a clause set $S(n)=\{C(p(1),\ldots,p(i),\ldots,p(n))\mid p(i)\in\{l_i, \sim l_i\}, i=1,\ldots, n\}$, then $S(n)$ is a contradiction that contains all the variables in $V$ and consists of $2^n \times n$ literals. $S(n)$ is called *the maximal contradiction generated by V.*

*Proof.* Based on the definition of $S(n)$, we know that there are $2^n$ literals that can be denoted as $C_1,\ldots, C_j,\ldots,C_{2^n}$. $S(n)$ can then be represented as CNF,

$$S(n) =\vee_{(x_1,\ldots,x_j,\ldots,x_{2^n})\in C_1\times\ldots\times C_j\times\ldots\times C_{2^n}, C_j\in S(n), j=1,\ldots,2^n} (x_1 \wedge\ldots\wedge x_j \wedge\ldots\wedge x_{2^n}).$$

Therefore, $S(n)$ is a contradiction if and only if $x_1 \wedge\ldots\wedge x_j \wedge\ldots\wedge x_{2^n}$ is unsatisfiable for any $(x_1,\ldots,x_j,\ldots,x_{2^n}) \in C_1 \times\ldots\times C_j \times\ldots\times C_{2^n}, C_j \in S(n), j = 1,\ldots,2^n$, i.e., there is complementary pair of literals among $x_1,\ldots,x_j,\ldots,x_{2^n}$ for any $(x_1,\ldots,x_j,\ldots,x_{2^n}) \in C_1 \times\ldots\times C_j \times\ldots\times C_{2^n}$, $C_j \in S(n), j = 1,\ldots,2^n$.

The following is proved by induction.

When $n= 1$, $S(1)=\{l_1, \sim l_1\}$, the conclusion holds.

When $n=2$, $S(2)=\{l_1 \vee l_2, l_1 \vee \sim l_2, \sim l_1 \vee l_2, \sim l_1 \vee \sim l_2\}$, the conclusion also holds.

Suppose that the conclusion holds when $n = k$, i.e., for any $(x_1,\ldots,x_j,\ldots,x_{2^k}) \in C_1 \times\ldots\times C_j \times\ldots\times C_{2^k}, C_j \in S(k), j = 1,\ldots,2^k$, there is complementary pair of literals among $x_1,\ldots,x_j,\ldots,x_{2^k}$.

When $n= k+1$, $S(k+1)=\{C(p(1),\ldots, p(i),\ldots, p(k), l_{k+1})\mid p(i) \in \{l_i, \sim l_i\}, i=1,\ldots, k\} \cup \{C(p(1),\ldots, p(i),\ldots, p(k), \sim l_{k+1})\mid p(i)\in\{l_i, \sim l_i\}, i=1,\ldots, k\}$.

If there exits $(x_{01},\ldots,x_{0i},\ldots,x_{02^k},y_{01},\ldots,y_{0j},\ldots,y_{02^k}) \in C_1 \times\ldots\times C_j \times\ldots\times C_{2^{k+1}}, C_j \in S(k+1), j = 1,\ldots,2^{k+1}$, such that there is no complementary pair of literals among $(x_{01},\ldots,x_{0i},\ldots,x_{02^k},y_{01},\ldots,y_{0j},\ldots,y_{02^k})$, where $C_r \in \{C(p(1),\ldots,p(i),\ldots,p(k), l_{k+1})\mid$ $p(i)\in\{l_i, \sim l_i\}$, $i$=1,…, $k\}$, $r = 1,\ldots,2^k$, $C_t \in\{C(p(1),\ldots, p(i),\ldots, p(k), \sim l_{k+1})| \ p(i)\in\{l_i, \sim l_i\}$, $i$ =1,…, $k\}$, $t = 2^k + 1,\ldots,2^{k+1}$, then the following two conclusions hold: (1) there is no complementary pair of literals among $x_{01},\ldots,x_{0i},\ldots,x_{02^k}$; and (2) there is no complementary pair of literals among $y_{01},\ldots,y_{0j},\ldots,y_{02^k}$.

It can be seen from (1) that, there exists $x_0 \in \{x_{01},\ldots,x_{0i},\ldots,x_{02^k}\}$ such that $x_0 = l_{k+1}$. If for any $x \in \{x_{01},\ldots,x_{0i},\ldots,x_{02^k}\}, x \neq l_{k+1}$, then $(x_{01},\ldots,x_{0j},\ldots,x_{02^k}) \in C_1 \times\ldots\times C_j \times\ldots\times C_{2^k}, C_j \in S(k)$. Based on the induction assumptions, there is complementary pair of literals among $x_{01},\ldots,x_{0i},\ldots,x_{02^k}$, which contradicts with (1). From (2), there exists $y_0 \in \{y_{01},\ldots,y_{0j},\ldots,y_{02^k}\}$ such that $y_0 = \sim l_{k+1}$. If for any $y \in \{y_{01},\ldots,y_{0j},\ldots,y_{02^k}\}$, $y \neq \sim l_{k+1}$, then $(y_{01},\ldots,y_{0j},\ldots,y_{02^k}) \in C_1 \times\ldots\times C_j \times\ldots\times C_{2^k}, C_j \in S(k)$. Based on the induction assumption, there is complementary pair of literals among $y_{01},\ldots,y_{0j},\ldots,y_{02^k}$, which contradicts with (2). Since $l_{k+1}$ and $\sim l_{k+1}$ is a complementary pair of literals, there is complementary pair of literals among $(x_{01},\ldots,x_{0i},\ldots,x_{02^k},y_{01},\ldots,y_{0j},\ldots,y_{02^k})$, and this contradicts to the assumption.

Therefore, for any $(x_1,\ldots,x_i,\ldots,x_{2^k},y_1,\ldots,y_j,\ldots,y_{2^k}) \in C_1 \times\ldots\times C_j \times\ldots\times C_{2^{k+1}}, C_j \in S(k+1)$, $j = 1,\ldots,2^{k+1}$, there is complementary pair of literals among $(x_1,\ldots,x_i,\ldots,x_{2^k},y_1,\ldots,y_j,\ldots,y_{2^k})$, which means that the conclusion holds when $n= k+1$.

Note that $S(n)=\{C(p(1),\ldots, p(i),\ldots, p(n))| \ p(i)\in\{l_i, \sim l_i\}, i=1,\ldots, n\}$ consists of all the $n$-ary clauses formed by the variables $\{l_1,\ldots, l_n\}$, where there are $2^n$ clauses and $2^n \times n$ literals. ■

For a given clause set $S$, its property can be only one of two cases: either $S$ is unsatisfiable or $S$ is satisfiable. Accordingly, the following discussion examines these two cases separately, analyzing the relationship between the maximal contradiction constructed from all propositional variables of $S$ and the property of $S$, and presenting methods for determining the property of $S$ based on this contradiction.

The relationship between the maximal contradiction of $S$ and the unsatisfiability of $S$ is given as follows:

**Theorem 4.2** Let $S=\{C_1, C_2,\ldots, C_m\}$ be a clause set in propositional logic, where the variable set of $S$ is $V(S) = \{l_1,\ldots, l_n\}$. For the clause set $S(n)=\{C(p(1),\ldots, p(i),\ldots, p(n))| \ p(i)\in\{l_i, \sim l_i\}, i=1,\ldots, n\}$, where clause $C(p(1),\ldots, p(i),\ldots, p(n))=\vee_{i=1}^{n} p(i)$, $p(i)\in\{l_i, \sim l_i\}$, $i=1,\ldots, n$, the following conclusions hold:

(1) For any $C_j \in S$, there exists $C(p(1),\ldots, p(i),\ldots, p(n))\in S(n)$, such that $C_j\subseteq C(p(1),\ldots, p(i),\ldots, p(n))$. It is called $C_j$ generates $C(p(1),\ldots, p(i),\ldots, p(n))$.

(2) For any $C_j\in S$, suppose $S(C_j)=\{C(p(1),\ldots, p(i),\ldots, p(n))\in S(n) \mid C_j\subseteq C(p(1),\ldots, p(i),\ldots, p(n))\}$, then $S$ is unsatisfiable if and only if $\cup_{i=1}^{m} S(C_j) = S(n)$.

*Proof.* (1) For any clause $C_j \in S$, suppose that $C_j = p_1 \vee \ldots \vee p_k \vee \ldots \vee p_t$, the for any $p_k$ ($k$ =1,…, $t$), there exists $i_k \in \{1, \ldots, n\}$ satisfying $p_k = i_k$ or $\sim i_k$. Therefore, there exists clause $C(p(1), \ldots, p(i), \ldots, p(n)) = p_1 \vee \ldots \vee p_k \vee \ldots \vee p_t \vee C^* \in S(n)$, and $C_j \subseteq C(p(1), \ldots, p(i), \ldots, p(n))$.

(2) **Sufficiency**. Apply disjointization processing to $\{S(C_j) \mid j = 1, \ldots, m\}$ first.

Let $S^*(C_1) = S(C_1)$,

$S^*(C_2) = S(C_2) - S^*(C_1)$,

$S^*(C_3) = S(C_3) - [S^*(C_1) \cup S^*(C_2)]$,

……

$S^*(C_j) = S(C_j) - \bigcup_{i=1,\ldots,j-1} S^*(C_i), j = 1, \ldots, m.$

Note that: (1) the elements of $\{S^*(C_j) \mid j=1, \ldots, m\}$ do not conclude each other, (2) the elements of $\{S^*(C_j) \mid j=1, \ldots, m\}$ are not empty and do not intersect each other, and (3) $\bigcup_{j=1,\ldots,m} S^*(C_j) = S(n)$. Therefore, suppose that the elements of $\{S^*(C_j) \mid j=1, \ldots, m\}$ are not empty, do not intersect each other, and the union of all the elements is $S(n)$.

There are $2^n$ clauses in $S(n)$, and can be denoted as $D_1, \ldots, D_i, \ldots, D_{2^n}$.

Suppose that $S(C_j) = \{D_{j_h} \in \{D_1, \cdots, D_i, \cdots, D_{2^n}\} \mid h = 1, \cdots, h_j\}, j = 1, \ldots, m$, and therefore $\sum_{j=1}^{m} h_j = 2^n$.

It is assumed that $D_{1_1}, \ldots, D_{1_{h_1}}, \ldots, D_{j_1}, \ldots, D_{j_{h_j}}, \ldots, D_{m_1}, \ldots, D_{m_{h_m}}$ correspond to $D_1, \ldots, D_i, \ldots, D_{2^n}$ one by one without loss of generality, and so, $\wedge_{j=1}^{m} C_j = \wedge_{j=1}^{m} [C_j \wedge \ldots \wedge C_j]$ where clause $C_j$ appears $h_j$ times. The Cartesian product of the corresponding set of literals of each clause of $\wedge_{j=1}^{m} C_j$ is $\prod_{j=1}^{m} [C_j \times \ldots \times C_j] \subseteq D_1 \times \cdots \times D_i \times \cdots \times D_{2^n}$ where clause $C_j$ appears $h_j$ times. It can be concluded that $\prod_{j=1}^{m} [C_j \times \ldots \times C_j]$ is unsatisfiable from the unsatisfiability of $D_1 \wedge \cdots \wedge D_i \wedge \cdots \wedge D_{2^n}$. Therefore, $S = \wedge_{j=1}^{m} C_j$ is unsatisfiable.

**Necessity**. Apply induction on $n$, the number of variables of $S$.

(1) When $n$=1, $S(n) = \{l_1, \sim l_1\}$. It can be seen from the unsatisfiability of $S = \{C_1, \ldots, C_m\}$ that $C_1 = l_1$, $C_2 = \sim l_1$ and so $S(C_1) = l_1$, $S(C_2) = \sim l_1$. Therefore, $\cup_{i=1}^{2} S(C_j) = \{l_1, \sim l_1\} = S(1)$, and the conclusion holds.

(2) Suppose that the conclusion holds when $n$=$k$, that is, $\cup_{i=1}^{m} S(C_j) = S(n)$ based on the unsatisfiability of $S = \{C_1, \ldots, C_m\}$.

(3) The following proves the conclusion holds when $n$=$k$+1.

As $S = \{C_1, \ldots, C_m\}$ is unsatisfiable, we can assume that: literals $l_{k+1}$ and $\sim l_{k+1}$ do not exist in clauses $C_1, \ldots, C_e$, literal $l_{k+1}$ exists but $\sim l_{k+1}$ does not exist in clauses $C_{e+1}, \ldots, C_f$, and literal $\sim l_{k+1}$ exists but $l_{k+1}$ does not exist in clauses $C_{f+1}, \ldots, C_m$. Therefore,

$$C_1 \wedge \ldots \wedge C_m = [C_1 \wedge \ldots \wedge C_e] \wedge [l_{k+1} \vee (C_{e+1}^* \wedge \ldots \wedge C_f^*)] \wedge [\sim l_{k+1} \vee (C_{f+1}^* \wedge \ldots \wedge C_m^*)]$$

$$= [C_1 \wedge \ldots \wedge C_e] \wedge \{[l_{k+1} \wedge \sim l_{k+1}] \vee [l_{k+1} \wedge (C_{f+1}^* \wedge \ldots \wedge C_m^*)] \vee [(C_{e+1}^* \wedge \ldots \wedge C_f^*) \wedge \sim l_{k+1}] \vee$$

$$[\,(C_{e+1}^{*}\wedge\ldots\wedge C_{f}^{*})\wedge(C_{f+1}^{*}\wedge\ldots\wedge C_{m}^{*})]\}$$

$$=\{[C_1\wedge\ldots\wedge C_e]\wedge[l_{k+1}\wedge\sim l_{k+1}]\}\vee$$

$$\{[C_1\wedge\ldots\wedge C_e]\wedge[l_{k+1}\wedge(C_{f+1}^{*}\wedge\ldots\wedge C_{m}^{*})]\}\vee$$

$$\{[C_1\wedge\ldots\wedge C_e]\wedge[\,(C_{e+1}^{*}\wedge\ldots\wedge C_{f}^{*})\wedge\sim l_{k+1}]\}\vee$$

$$\{[C_1\wedge\ldots\wedge C_e]\wedge[\,(C_{e+1}^{*}\wedge\ldots\wedge C_{f}^{*})\wedge(C_{f+1}^{*}\wedge\ldots\wedge C_{m}^{*})]\}$$

$$=\{[C_1\wedge\ldots\wedge C_e]\wedge[l_{k+1}\wedge(C_{f+1}^{*}\wedge\ldots\wedge C_{m}^{*})]\}\vee$$

$$\{[C_1\wedge\ldots\wedge C_e]\wedge[\,(C_{e+1}^{*}\wedge\ldots\wedge C_{f}^{*})\wedge\sim l_{k+1}]\}\vee$$

$$\{[C_1\wedge\ldots\wedge C_e]\wedge[\,(C_{e+1}^{*}\wedge\ldots\wedge C_{f}^{*})\wedge(C_{f+1}^{*}\wedge\ldots\wedge C_{m}^{*})]\}.$$

The following three equations are unsatisfiable follows from the unsatisfiability of $C_1\wedge\ldots\wedge C_m$.

$$[C_1\wedge\ldots\wedge C_e]\wedge[l_{k+1}\wedge(C_{f+1}^{*}\wedge\ldots\wedge C_{m}^{*})],$$

$$[C_1\wedge\ldots\wedge C_e]\wedge[(C_{e+1}^{*}\wedge\ldots\wedge C_{f}^{*})\wedge\sim l_{k+1}],$$

$$[C_1\wedge\ldots\wedge C_e]\wedge[(C_{e+1}^{*}\wedge\ldots\wedge C_{f}^{*})\wedge(C_{f+1}^{*}\wedge\ldots\wedge C_{m}^{*})].$$

Since both the literals $l_{k+1}$ and $\sim l_{k+1}$ do not exist in clauses $C_1,\ldots, C_e, C_{f+1}^{*},\ldots,C_m^{*}$, it can be obtained that $C_1\wedge\ldots\wedge C_e\wedge C_{f+1}^{*}\wedge\ldots\wedge C_m^{*}$ is unsatisfiable from unsatisfiability of $C_1\wedge\ldots\wedge C_e\wedge l_{k+1}\wedge C_{f+1}^{*}\wedge\ldots\wedge C_m^{*}$, and the number of variables of $C_1\wedge\ldots\wedge C_e\wedge C_{f+1}^{*}\wedge\ldots\wedge C_m^{*}$ being $n\le k$.

If $n<k$, suppose the variables in $C_1\wedge\ldots\wedge C_e\wedge C_{f+1}^{*}\wedge\ldots\wedge C_m^{*}$ are $x_1,\ldots, x_n$, then $C_1\wedge\ldots\wedge C_e\wedge C_{f+1}^{*}\wedge\ldots\wedge C_m^{*}$ can be transformed to $C_1\wedge\ldots\wedge C_e\wedge C_{f+1}^{*}\wedge\ldots\wedge C_m^{*}\wedge x_{n+1}\wedge\ldots\wedge x_k$, and so there are $k$ variables in $C_1\wedge\ldots\wedge C_e\wedge C_{f+1}^{*}\wedge\ldots\wedge C_m^{*}\wedge x_{n+1}\wedge\ldots\wedge x_k$, and $C_1\wedge\ldots\wedge C_e\wedge C_{f+1}^{*}\wedge\ldots\wedge C_m^{*}\wedge x_{n+1}\wedge\ldots\wedge x_k$ is unsatisfiable if and only if $C_1\wedge\ldots\wedge C_e\wedge C_{f+1}^{*}\wedge\ldots\wedge C_m^{*}$ is unsatisfiable. Therefore, suppose that there are $k$ variables in $C_1\wedge\ldots\wedge C_e\wedge C_{f+1}^{*}\wedge\ldots\wedge C_m^{*}$. Based on the induction assumption, $\cup_{r=1}^{e} S(C_r)\cup\left(\cup_{q=f+1}^{m} S(C_q^{*})\right)=S(k)$. There are $2^k$ clauses in $S(k)$, and $k$ variables in each clause. Add literal $\sim l_{k+1}$ to each clause in $S(k)$, then for any $r\in\{1,\ldots, e\}$, $S(C_r)$ becomes $\{C\vee\sim l_{k+1}|C\in S(C_r)\}$. Denote $S(C_r)^1=\{C\vee\sim l_{k+1}|C\in S(C_r)\}$, where there are $k$+1 variables in each clause. For any $q\in\{f+1,\ldots, m\}$, add literal $l_{k+1}$ to each clause of $S(C_q^{*})$, which makes $S(C_q^{*})$ becoming $S(C_q)$, where there are $k$+1 variables in each clause. Hence, $\cup_{r=1}^{e} S(C_r)\cup(\cup_{q=f+1}^{m} S(C_q^{*}))$ becomes $\cup_{r=1}^{e} S(C_r)^1\cup(\cup_{q=f+1}^{m} S(C_q))$, where there are $2^k$ clauses in $\cup_{r=1}^{e} S(C_r)^1\cup(\cup_{q=f+1}^{m} S(C_q))$, are $k$+1 variables in each clause.

Similarly, as both the literals $l_{k+1}$ and $\sim l_{k+1}$ do not exist in clauses $C_1,\ldots,C_e, C_{e+1}^{*},\ldots,C_f^{*}$, it can be obtained that $C_1\wedge\ldots\wedge C_e\wedge C_{e+1}^{*}\wedge\ldots\wedge C_f^{*}$ is unsatisfiable from unsatisfiability of $C_1\wedge\ldots\wedge C_e\wedge\sim l_{k+1}\wedge C_{e+1}^{*}\wedge\ldots\wedge C_f^{*}$. Suppose that there are $k$ variables in $C_1\wedge\ldots\wedge C_e\wedge C_{e+1}^{*}\wedge\ldots\wedge C_f^{*}$. It follows from the induction assumption that $\cup_{r=1}^{e} S(C_r)\cup$ $\left(\cup_{p=e+1}^{f} S(C_p^*)\right) = S(k)$, where there are $2^k$ clauses in $S(k)$, and $k$ variables in each clause. Add literal $l_{k+1}$ to each clause in $S(k)$, then for any $r\in\{1,\ldots, e\}$, $S(C_r)$ becomes $\{C \vee l_{k+1}|C\in S(C_r)\}$. Denote $S(C_r)^2 = \{C \vee l_{k+1}|C \in S(C_r)\}$, where there are $k+1$ variables in each clause. For any $p\in\{e+1,\ldots,f\}$, add literal $l_{k+1}$ to each clause of $S(C_p^*)$, which makes $S(C_p^*)$ becoming $S(C_p)$, where there are $k+1$ variables in each clause. Hence, $\cup_{r=1}^{e} S(C_r) \cup (\cup_{p=e+1}^{m} S(C_p^*))$ becomes $\cup_{r=1}^{e} S(C_r)^2 \cup (\cup_{p=e+1}^{m} S(C_p))$, where there are $2^k$ clauses in $\cup_{r=1}^{e} S(C_r)^2 \cup (\cup_{p=e+1}^{m} S(C_p))$, are $k+1$ variables in each clause.

Take the union of $\cup_{r=1}^{e} S(C_r)^1$, $(\cup_{q=f+1}^{m} S(C_q))$,$\cup_{r=1}^{e} S(C_r)^2$, $(\cup_{p=e+1}^{f} S(C_p))$ as they do not intersect with each other to obtain $2^k \times 2$ clauses, where there are $k+1$ variables in each clause. $S(C_r)^1 \cup S(C_r)^2$ contains all the clauses generated by $C_r$. Therefore,

$$\cup_{r=1}^{e} (S(C_r)^1 \cup S(C_r)^2) \cup \left(\cup_{q=f+1}^{m} S(C_q)\right) \cup \left(\cup_{P=e+1}^{f} S(C_p)\right) = \cup_{j=1}^{m} S(C_j),$$

which means that the conclusion holds for $n=k+1$.

Sum up above, the conclusion holds. ■

It can be seen from the proof of Theorem 4.2 that redundant clauses are allowed in the original clause set $S =\{C_1, C_2,\ldots, C_m\}$. Accordingly, a method for determining the unsatisfiability of a clause set by utilizing the maximal contradiction deduction can also be obtained based on Theorem 4.2, as illustrated firstly in the following example.

**Example 4.1** Suppose that $S = \{C_1, C_2, C_3, C_4, C_5, C_6\}$ is a clause set in propositional logic, with $C_1 = l_1 \vee l_2$, $C_2 = l_2 \vee l_3$, $C_3 = l_3 \vee l_4$, $C_4 = \sim l_3 \vee \sim l_1$, $C_5 = \sim l_4 \vee \sim l_2$, $C_6 = \sim l_2 \vee \sim l_3$, where $l_1$, $l_2$, $l_3$, $l_4$ are propositional variables. Please check the unsatisfiability of $S$ using the maximal contradiction deduction.

**Method 1**.

*Step 1*. Construct the maximal contradiction $S(3)$ based on variables $l_1$, $l_2$, $l_3$ as follows.

$$l_1 \vee l_2 \vee l_3$$
$$l_1 \vee l_2 \vee \sim l_3$$
$$l_1 \vee \sim l_2 \vee l_3$$
$$l_1 \vee \sim l_2 \vee \sim l_3$$
$$\sim l_1 \vee l_2 \vee l_3$$
$$\sim l_1 \vee l_2 \vee \sim l_3$$
$$\sim l_1 \vee \sim l_2 \vee l_3$$
$$\sim l_1 \vee \sim l_2 \vee \sim l_3$$

*Step 2*. Generate a subset of $S(3)$ based on clauses $C_1$, $C_2$, $C_3$, $C_4$, $C_6$, and obtain $R_0 = l_4$.

*Step 3*. Put $R_0 = l_4$ into $S$, and apply unit clause rule to obtain $S_1$:

$$C_1 = l_1 \vee l_2$$

$$C_2 = l_2 \vee l_3$$

$$C_4 = \sim l_3 \vee \sim l_1$$

$$C_5 = \sim l_2$$

$$C_6 = \sim l_2 \vee \sim l_3$$

*Step 4*. Generate a subset of $S(3)$ based on $S_1$, and it can be drawn that $S_1$ is unsatisfiable, and therefore $S$ is unsatisfiable.

**Method 2**.

*Step 1*. Construct the maximal contradiction $S(2)$ based on variables $l_2$, $l_3$ as follows.

$$l_2 \vee l_3$$

$$l_2 \vee \sim l_3$$

$$\sim l_2 \vee l_3$$

$$\sim l_2 \vee \sim l_3$$

*Step 2*. Generate a subset of $S(2)$ based on clauses $C_1 = l_1 \vee l_2$, $C_2 = l_2 \vee l_3$, $C_6 = \sim l_2 \vee \sim l_3$, $C_5 = \sim l_4 \vee \sim l_2$, and obtain $C_7 = l_1 \vee \sim l_4$. Put $C_7$ into $S$.

*Step 3*. Construct the maximal contradiction $S(3)^*$ based on variables $l_1$, $l_3$, $l_4$ as follows:

$$l_1 \vee l_3 \vee l_4$$

$$l_1 \vee l_3 \vee \sim l_4$$

$$l_1 \vee \sim l_3 \vee l_4$$

$$l_1 \vee \sim l_3 \vee \sim l_4$$

$$\sim l_1 \vee l_3 \vee l_4$$

$$\sim l_1 \vee l_3 \vee \sim l_4$$

$$\sim l_1 \vee \sim l_3 \vee l_4$$

$$\sim l_1 \vee \sim l_3 \vee \sim l_4$$

And generate a subset of $S(3)^*$ based on clauses $C_3$, $C_4$, $C_5$, $C_6$, $C_7$, and obtain $C_8 = \sim l_2$.

*Step 4*. Put $C_1$ — $C_8$ together to obtain an empty set.

- **Procedure for Determining the Unsatisfiability of a Propositional Clause Set via Maximal Contradiction Deduction**

The following outlines the steps for determining the unsatisfiability of a propositional logic clause set $S$ using the deduction of the maximal contradiction:

**Step 0**. Perform direct redundancy processing on $S$ to obtain $S_0$.

**Step 1**. Construct the maximal contradiction $C(V_0)$ generated by some variables of $S_0$ (denote the set of these variables as $V_0$). Let $L(V_0)$ be the complete set of literals produced by $V_0$, and define $C(V_0) = \{O_1(V_0), \ldots, O_t(V_0), \ldots, O_{2^{|V_0|}}(V_0)\}$. For each $t \in \{1, \ldots, 2^{|V_0|}\}$, $O_t(V_0)$ is a clause generated by $V_0$, with $|O_t(V_0)| = |V_0|$, meaning $O_t(V_0)$ contains $|V_0|$ distinct literals from

$L(V_0)$, and $O_t(V_0)$ is not a tautology. The set of literals in $O_t(V_0)$ is still denoted as $O_t(V_0)$. Proceed to Step 2.

**Step 2**. Select clauses $C_1,\ldots, C_t,\ldots,C_{2^{|V_0|}}$ consequently satisfying:

(1) For $O_1(V_0)$, select clause $C_1$ that satisfies the following two conditions:

$C_1^1$: $C_1 \cap O_1(V_0) \neq \emptyset$;

$C_1^2$: $|C_1 - O_1(V_0)|$ is minimum in $S_0$.

(2) For $O_2(V_0)$, select clause $C_2$ that satisfies the following two conditions:

$C_2^1$: $C_2 \cap O_2(V_0) \neq \emptyset$;

$C_2^2$: $|[C_2 - O_2(V_0)] \cap [C_1 - O_1(V_0)]|$ is maximum, which is to guarantee the number of the remaining literals that are not the same to be minimum.

$C_2^3$: $|C_2 - [O_2(V_0) \cup (C_1 - O_1(V_0))]|$ is required to be minimum in $S_0$ under the premise of satisfying $C_2^1$ and $C_2^2$.

……

($t$) For $O_t(V_0)$, select clause $C_t$ that satisfies the following two conditions:

$C_t^1$: $C_t \cap O_t(V_0) \neq \emptyset$;

$C_t^2$: $[C_t - O_t(V_0)] \cap \cap_{i=1}^{t-1} [C_i - O_i(V_0)]$ is maximum.

$C_t^3$: $|C_t - [O_t(V_0) \cup \cup_{i=1}^{t-1} (C_i - O_i(V_0))]|$ is required to be minimum in $S_0$ under the premise of satisfying $C_t^1$ and $C_t^2$.

……

($2^{|V_0|}$) For $O_{2^{|V_0|}}(V_0)$, select clause $C_{2^{|V_0|}}$ that satisfies the following two conditions:

$C_{2^{|V_0|}}{}^1$: $C_{2^{|V_0|}} \mid O_{2^{|V_0|}}(V_0) \neq \emptyset$;

$C_{2^{|V_0|}}{}^2$: $|[C_{2^{|V_0|}} - O_{2^{|V_0|}}(V_0)] \cap \cap_{i=1}^{2^{|V_0|}-1} [C_i - O_i(V_0)]|$ is maximum;

$C_{2^{|V_0|}}{}^3$: $|C_{2^{|V_0|}} - [O_{2^{|V_0|}}(V_0) \cup \cup_{i=1}^{2^{|V_0|}-1} (C_i - O_i(V_0))]|$ is required to be minimum in $S_0$ under the premise of satisfying $C_{2^{|V_0|}}{}^1$ and $C_{2^{|V_0|}}{}^2$.

Here, $C_1,\ldots, C_t,\ldots,C_{2^{|V_0|}}$ can be repeated and the result $\vee_{t\in\{1,\ldots,2^{|V_0|}\}} (C_t - O_t(V_0))$ is obtained and denoted as $R_0$. Proceed to Step 3.

According to this process, $\prod_{t\in\{1,\ldots,2^{|V_0|}\}}[C_t \cap O_t(V_0)] \subseteq \prod_{t\in\{1,\ldots,2^{|V_0|}\}} O_t(V_0)$ can be obtained. Note that $\prod_{t\in\{1,\ldots,2^{|V_0|}\}} O_t(V_0)$ is a contradiction, and therefore $\prod_{t\in\{1,\ldots,2^{|V_0|}\}}[C_t \cap O_t(V_0)]$ is also a contradiction and the remaining literals are very few.

**Step 3**. Perform direct redundancy processing on $S_0 \cup \{R_0\}$, denote the resulting set as $S_1$, and proceed to Step 4.

**Step 4**. Repeat Step 1 for $S_1$.

……

until the empty clause is obtained or the problem cannot be determined.

In the following, we examine the relationship between the maximal contradiction and the satisfiability of a clause set, and present a method for determining satisfiability using the maximal contradiction.

**Definition 4.1** Let $S=\{C_1, C_2,\ldots, C_m\}$ be a clause set in propositional logic, where the variable set of $S$ is $V(S)$. Clause $D$ is called a maximal clause regarding $V(S)$, if for any $x\in V(S)$, exactly one of $x$ and $\sim x$ is present in clause $D$, and $D$ contains $|V(S)|$ literals.

**Definition 4.2** Let $S=\{C_1, C_2,\ldots, C_m\}$ be a clause set in propositional logic, where the variable set of $S$ is $V(S)$, and $D$ is a maximal clause regarding $V(S)$. If there is no clause $C$ in $S$ such that $C\subseteq D$, then $D$ is said to be non-expandable by $S$; that is, $D\in S(n) -\cup_{j=1}^{m} S(C_j)$, where $S(n)$ is the maximal contradiction generated by $V(S)$.

**Theorem 4.3** Let $S=\{C_1, C_2,\ldots, C_m\}$ be a clause set in propositional logic, where the variable set of $S$ is $V(S) = \{l_1,\ldots, l_n\}$. $S(n)$ is the maximal contradiction generated by $V(S)$, then the following conclusions are equivalent.

(1) $S$ is satisfiable;

(2) $S(n) -\cup_{j=1}^{m} S(C_j) \neq\phi$;

(3) There is maximal clause $D$ regarding variable set $V(S)$ that is non- expandable by $S$;

(4) There is maximal clause $D$ regarding variable set $V(S)$, such that for any $C\in S$, there exists $x \in C$, $\sim x \in D$;

(5) There is maximal clause $D$ regarding variable set $V(S)$, such that for any $C\in S$, $C\cap\sim D\neq \phi$, where $\sim D$ is the clause obtained by negating all literals in $D$.

*Proof.* The equivalence of (1) and (2):

Since $S$ is unsatisfiable if and only if $\cup_{j=1}^{m} S(C_j) = S(n)$, and $\cup_{j=1}^{m} S(C_j) \subseteq S(n)$, it follows that $S$ is satisfiable if and only if $\cup_{j=1}^{m} S(C_j) \subset S(n)$, i.e., $S(n) -\cup_{j=1}^{m} S(C_j) \neq \emptyset$.

The equivalence of (2) and (3) is obvious.

The equivalence of (3) and (4):

(**Necessity**) If there exists $C_{i_0} \in S$ such that for every $x \in C_{i_0}$, $\sim x \notin D$, then since $D$ is a maximal clause with respect to the variable set $V(S)$, it follows that $x \in D$. Therefore, $C_{i_0} \subseteq D$, which contradicts the fact that $D$ is non-expandable by $S$. Hence, the conclusion holds.

(**Sufficiency**) If $D$ can be expanded by $S$, then there exists $C_{i_0} \in S$ such that $C_{i_0} \subseteq D$. Thus, for every $x \in C_{i_0}$, $x \in D$, and $\sim x \notin D$. This contradicts (4), so the assumption is false, and conclusion (3) holds.

The equivalence of (4) and (5):

It suffices to note the construction of $\sim D$, i.e., $\sim D = \{\sim y | y \in D\}$, which means $y \in D$ if and only if $\sim y \in \sim D$. ■

According to conclusion (2) of Theorem 4.3, the satisfiability of a clause set can be determined using the maximal contradiction, as shown in Example 4.2.

**Example 4.2** Let $S$={$C_1$, $C_2$, $C_3$, $C_4$} be a clause set in propositional logic, with $C_1 = \sim l_1$, $C_2 = \sim l_3 \vee l_4$, $C_3 = \sim l_2 \vee \sim l_4$, $C_4 = l_1 \vee l_2 \vee l_3$, where $l_1$, $l_2$, $l_3$, $l_4$ are propositional variables. Please check the satisfiability of $S$ using the maximal contradiction.

Actually, all the variables in $S$ are $l_1$, $l_2$, $l_3$, $l_4$, and the maximal contradiction generated by $l_1$, $l_2$, $l_3$, $l_4$ is $S(4)$:

$$G_1 = l_1 \vee l_2 \vee l_3 \vee l_4$$

$$G_2 = l_1 \vee l_2 \vee l_3 \vee \sim l_4$$

$$G_3 = l_1 \vee l_2 \vee \sim l_3 \vee l_4$$

$$\boldsymbol{G_4 = l_1 \vee l_2 \vee \sim l_3 \vee \sim l_4}$$

$$\boldsymbol{G_5 = l_1 \vee \sim l_2 \vee l_3 \vee l_4}$$

$$G_6 = l_1 \vee \sim l_2 \vee l_3 \vee \sim l_4$$

$$G_7 = l_1 \vee \sim l_2 \vee \sim l_3 \vee l_4$$

$$G_8 = l_1 \vee \sim l_2 \vee \sim l_3 \vee \sim l_4$$

$$G_9 = \sim l_1 \vee l_2 \vee l_3 \vee l_4$$

$$G_{10} = \sim l_1 \vee l_2 \vee l_3 \vee \sim l_4$$

$$G_{11} = \sim l_1 \vee l_2 \vee \sim l_3 \vee l_4$$

$$G_{12} = \sim l_1 \vee l_2 \vee \sim l_3 \vee \sim l_4$$

$$G_{13} = \sim l_1 \vee \sim l_2 \vee l_3 \vee l_4$$

$$G_{14} = \sim l_1 \vee \sim l_2 \vee l_3 \vee \sim l_4$$

$$G_{15} = \sim l_1 \vee \sim l_2 \vee \sim l_3 \vee l_4$$

$$G_{16} = \sim l_1 \vee \sim l_2 \vee \sim l_3 \vee \sim l_4$$

It is obvious that $G_4$ and $G_5$ cannot be generated by any clause of $S$, that is, $S(4) - \cup_{j=1}^{4} S(C_j) \neq \emptyset$, and so $S$ is satisfiable.

**Theorem 4.4** Let $S = \{C_1, \cdots, C_m\}$ be a satisfiable clause set in propositional logic, $S(n)$ is the maximal contradiction generated by all propositional variables appearing in $S$, and $S(n) - \left(\cup_{j=1}^{m} S(C_j)\right) = \{D_1, \cdots, D_t\}$. Then, for any $(x_1, \ldots, x_m) \in \cup_{j=1}^{t} \prod_{i=1}^{m} (C_i \cap \sim D_j)$, the set $\{x_1, \ldots, x_m\}$ is a satisfiable instance of $S$, where $\sim D_j$ is the clause obtained by negating all literals in $D_j$, $j = 1, \ldots, m$

*Part 1 of Proof.* Apply induction on $m$, the number of clauses of $S$.

(1) When $m$=1, it is obvious that $S$ is satisfiable as there is only one clause in $S$. For any $x \in \cup_{j=1}^{t} (C_1 \cap \sim D_j)$, $x$ is a satisfiable instance of $S = \{C_1\}$ as $x \in C_1$, and so the conclusion holds.

(2) Assume that the conclusion holds when $m$=$k$, that is, for any $(x_1, \dots, x_k) \in \cup_{j=1}^{t} \prod_{i=1}^{k} (C_i \cap\sim D_j)$, $\{x_1, \dots, x_k\}$ is a satisfiable instance of $S = \{C_1, \cdots, C_k\}$.

(3) When $m$=$k$+1, suppose that $S = \{C_1, \cdots, C_k, C_{k+1}\}$ is satisfiable, and so is $S = \{C_1, \cdots, C_k\}$, then the induction assumption can be applied to $\{C_1, \cdots, C_k\}$. Let $S(n) - \left(\cup_{j=1}^{k+1} S(C_j)\right) = \{D_1, \cdots, D_e\}$.

For every $D \in \{D_1, \cdots, D_e\}$, it is known from the induction assumption that for any $(x_1, \dots, x_k) \in \prod_{i=1}^{k} (C_i \cap\sim D)$, $\{x_1, \dots, x_k\}$ is a satisfiable instance of $S = \{C_1, \cdots, C_k\}$, i.e., there is no complementary pair of literals among $\{x_1, \dots, x_k\}$. We prove that there is no complementary pair of literals among $\{x_1, \dots, x_k, x_{k+1}\}$, for any $(x_1, \dots, x_k) \in \prod_{i=1}^{k} (C_i \cap\sim D)$ and $x_{k+1} \in C_{k+1} \cap\sim D$.

In fact, it can be seen that $x_i \in\sim D$ as $x_i \in C_i \cap\sim D$, i.e., $\sim x_i \in D$, $i$=1,…, $k$. From $x_{k+1} \in C_{k+1} \cap\sim D$, we know that $x_{k+1} \in\sim D$, i.e., $\sim x_{k+1} \in D$. Since $D$ is a maximal clause regarding the variable set $V(S)$, there is no complementary pair of literals in $D$. Therefore, there is no complementary pair of literals among $\{\sim x_1, \dots, \sim x_k, \sim x_{k+1}\}$, and there is no complementary pair of literals among $\{x_1, \dots, x_k, x_{k+1}\}$ as well.

Therefore, for any $(x_1, \dots, x_k, x_{k+1}) \in \cup_{j=1}^{e} \prod_{i=1}^{k+1} (C_i \cap\sim D_j)$, there is no complementary pair of literals among $\{x_1, \dots, x_k, x_{k+1}\}$, and so $\{x_1, \dots, x_k, x_{k+1}\}$ is a satisfiable instance of $S = \{C_1, \dots, C_k, C_{k+1}\}$.

*Part 2 of Proof.*

(1) For any $C \in S = \{C_1, \dots, C_m\}$, for any $D \in S(n) - \cup_{j=1}^{m} S(C_j) = \{D_1, \dots, D_t\}$, $C \cap \sim D \neq \emptyset$. In fact, for any $C \in S$, for any $D \in \{D_1, \dots, D_t\}$, as $C \not\subseteq D$, there exists $x_0 \in C$, $x_0 \notin D$, since $D$ is a maximal clause, then $\sim x_0 \in D$, i.e., $x_0 \in \sim D$, and therefore $x_0 \in C \cap \sim D \neq \emptyset$, and further $(C_1 \cap \sim D) \times \dots \times (C_m \cap \sim D) \neq \emptyset$.

(2) For any $D \in \{D_1, \dots, D_t\}$, for any $\{x_1, \dots, x_m\} \in (C_1 \cap \sim D) \times \dots \times (C_m \cap \sim D)$, there is no complementary pair of literals among $\{x_1, \dots, x_m\}$. Actually, if there exist $x_{i_0}$ and $x_{j_0}$ that are complementary to each other, it means that there is complementary pair of literals in $D$, as both $x_{i_0}$ and $x_{j_0}$ are in $D$, which contradicts that $D$ is a maximal clause. Hence, there is no complementary pair of literals among $\{x_1, \dots, x_m\}$, and $\{x_1, \dots, x_m\}$ is a satisfiable instance of $S$.

It can be seen from the arbitrariness of $D$ that for any $(x_1, \dots, x_m) \in \cup_{j=1}^{t} \prod_{i=1}^{m} (C_i \cap\sim D_j)$, $\{x_1, \dots, x_m\}$ is a satisfiable instance of $S$. ■

- **Procedure (Method I) to find a satisfiable instance**

From Theorem 4.4, we know that the maximal contradiction can not only be applied to determine the satisfiability of a clause set but also provide a satisfiable instance when the clause set is satisfiable. The specific method is outlined as follows: Let $S = \{C_1, \cdots, C_m\}$ be a satisfiable clause set in propositional logic, and let $V(S)$ be the set of variables in $S$. Suppose $|V(S)| = n$, and let $S(n)$ be the maximal contradiction generated by $V(S)$. According to Theorem 4.4, $S(n) - \cup_{j=1}^{m} S(C_j) \neq \emptyset$. The specific steps to find a satisfiable instance are summarized:

**Step 1**. Generate a candidate maximal clause $D$ with respect to the variable set $V(S)$ randomly, and turn to Step 2.

**Step 2**. Check whether $D \in S(n) - \cup_{j=1}^{m} S(C_j)$ holds or not. If it holds, proceed to Step 3; otherwise, return to Step 1.

**Step 3**. Output $\sim D$, and proceed to Step 4.

**Step 4**. Select $x_i \in C_i \cap \sim D$, $i=1,\ldots, m$, consequently, then $\{x_1, \ldots, x_m\}$ is a satisfiable instance of $S$.

The following example provides an illustration:

**Example 4.3** Let $S=\{C_1, C_2, C_3\}$ be a clause set in propositional logic, with $C_1 = l_1 \vee l_2$, $C_2 = l_2 \vee l_3$, $C_3 = \sim l_2 \vee \sim l_3$, where $l_1, l_2, l_3$ are propositional variables. Please check the satisfiability of $S$ using the maximal contradiction, and provides a satisfiability instance of $S$.

*Step 1*. Generate a maximal clause $S(l_1, l_2, l_3)$ base on $l_1, l_2, l_3$ appearing in $S$ as follows.

$$l_1 \vee l_2 \vee l_3$$
$$l_1 \vee l_2 \vee \sim l_3$$
$$l_1 \vee \sim l_2 \vee l_3$$
$$l_1 \vee \sim l_2 \vee \sim l_3$$
$$\sim l_1 \vee l_2 \vee l_3$$
$$\sim l_1 \vee l_2 \vee \sim l_3$$
$$\sim l_1 \vee \sim l_2 \vee l_3$$
$$\sim l_1 \vee \sim l_2 \vee \sim l_3$$

*Step 2*. For the clause $l_1 \vee \sim l_2 \vee l_3$ in $S(l_1, l_2, l_3)$, it is evident that the literal $l_2$ from clause $C_1$, the literal $l_2$ from clause $C_2$, and the literal $\sim l_3$ from clause $C_3$ do not appear in the clause $l_1 \vee \sim l_2 \vee l_3$. Therefore, the clause set $S$ is satisfiable. Let $D = l_1 \vee \sim l_2 \vee l_3$ for convenience.

*Step 3*. Denote the clause obtained by negating every literal in $D$ as $\overline{D}$, i.e., $\overline{D} = \sim l_1 \vee l_2 \vee \sim l_3$. Clearly, the literal $l_2$ from clause $C_1$, the literal $l_2$ from clause $C_2$, and the literal $\sim l_3$ from clause $C_3$ all appear in $\overline{D}$. Thus, the literal $l_2$ from clause $C_1$, the literal $l_2$ from clause $C_2$, and the literal $\sim l_3$ from clause $C_3$ collectively form a satisfiable instance of $S$.

The following corollary can be obtained from Theorem 4.4.

**Corollary 4.1** Let $S = \{C_1, \cdots, C_m\}$ be a satisfiable clause set in propositional logic, $V(S)=\{1,\ldots, n\}$ be the set of variables in $S$, and $S(n)$ be the maximal contradiction generated by $V(S)$, then the following conclusions hold.

(1) If all clauses in $S$ are maximal clauses, then $S$ is unsatisfiable if and only if $2^n = m$.

(2) For any $D \in S(n) - \cup_{j=1}^{m} S(C_j)$, let $I_D(j) = \begin{cases} 0, j \in D \\ 1, j \notin D \end{cases} = \begin{cases} 0, j \in D \\ 1, \bar{j} \in D \end{cases}$, $j = 1, \ldots, n$, then $I_D$ is a satisfiability instance of $S$.

*Proof.* (1) Since all the clauses of $S$ are maximal clauses, i.e., for any $C \in S$, $S(C) = \{C\}$, and so $\cup_{j=1}^{m} S(C_j) = \{C_1, \ldots, C_m\} = S$, that is, $S(n) - \cup_{j=1}^{m} S(C_j) = S(n) - \{C_1, \ldots, C_m\}$. Because $S$ is unsatisfiable if and only if $S(n) - \cup_{j=1}^{m} S(C_j) = \emptyset$, then $S(n) - \{C_1, \ldots, C_m\} = \emptyset$, that is, $S(n) = \{C_1, \ldots, C_m\}$, $2^n = m$. Therefore, the conclusion holds.

(2) Since for any $D \in S(n) - \cup_{j=1}^{m} S(C_j)$, $\{x_1, \ldots, x_m\}$ is a satisfiable instance of $S = \{C_1, \ldots, C_m\}$ for any $(x_1, \ldots, x_m) \in \prod_{i=1}^{m} (C_i \cap \sim D)$, we know that $\sim D$ is a satisfiable instance of $S$, i.e., $\sim D(j) = \begin{cases} 0, j \notin \sim D \\ 1, j \in \sim D \end{cases} = \begin{cases} 0, \sim j \in \sim D \\ 1, j \in \sim D \end{cases}$. Because both $D$ and $\sim D$ are maximal clauses regarding $V(S)$, we know that $j \in D$ if and only if $j \notin \sim D$ if and only if $\sim j \in \sim D$. Therefore, $I_D(j) = \sim D(j)$, $j = 1, \ldots, n$, i.e., $I_D$ is a satisfiable instance of $S$. ■

- **Improved method II to find a satisfiable instance**

According to (2) of Corollary 4.1, to find a satisfiable instance of $S$, it is only necessary to find $D \in S(n) - \cup_{j=1}^{m} S(C_j)$. Thus, the following improved method for finding a satisfiable instance can be obtained.

Let $S = \{C_1, \ldots, C_m\}$ be a satisfiable clause set in propositional logic, and $V(S) = \{1, \ldots, n\}$ be the set of variables of $S$. Let $K$ be the number of times set for constructing maximal clauses with respect to $V(S)$. The specific steps of the simplified method for finding a satisfiable instance are summarized as follows:

**Step 1**. If the number of times $k \leq K$ for randomly (or by other methods) generating a maximal clause about $V(S)$, then randomly (or by other methods) generate a maximal clause regarding $V(S)$ as $D(k) = p(1) \vee \ldots \vee p(i) \vee \ldots \vee p(n)$, where $p(i) \in \{i, \sim i\}$, $i = 1, \cdots, n$, and proceed to Step 2; if the number of times $k > K$ for randomly (or by given methods) generating a maximal clause about $V(S)$, then stop.

**Step 2.** Delete the clauses containing $\sim p(1)$ from $S$ to obtain the clause set $S_1$. If $S_1 = \emptyset$, then $I_D$ is a satisfiability instance of $S$; if $S_1 \neq \emptyset$: when $n > 1$, proceed to Step 3; when $n = 1$, return to Step 1.

**Step 3**. Repeat Step 2 for $\sim p(2)$ and $S_1$.

**Remark 4.1** Within the preset limit $K$ for constructing maximal clauses about $V(S)$, if a maximal clause about $V(S)$, i.e., $D(k) = p(1) \vee \ldots \vee p(i) \vee \ldots \vee p(n)$ is randomly generated, and by sequentially deleting clauses containing $\sim p(1)$, $\sim p(2)$, …, $\sim p(t)$ $(t \leq n)$ from $S = \{C_1, \ldots,$ $C_m\}$, which indicates that the sequentially deleted clauses cannot be expanded to $D(k)$. In fact, if $C$ contains $\sim p(i)$, since $p(i)\in D(k)$, it follows that $\sim p(i) \notin D(k)$. Thus, $C$ cannot be contained in $D(k)$, meaning $C$ cannot be expanded to a clause of $D(k)$ so as to obtain an empty set, then it means $D(k)\in S(n) - \cup_{j=1}^{m} S(C_j) \neq \emptyset$. According to the conclusion (2) of Corollary 4.1, a satisfiability instance $I_{D(k)}$ is obtained.

The following example provides an illustration:

**Example 4.4** Let $S=\{C_1, C_2, C_3\}$ be a clause set in propositional logic, with $C_1 = l_1 \vee l_2$, $C_2 = \sim l_2 \vee l_3$, $C_3 = \sim l_1 \vee \sim l_3$, where $l_1$, $l_2$, $l_3$ are propositional variables. Please check the satisfiability of $S$ using the maximal contradiction, and provides a satisfiability instance of $S$.

*Step 1*. Construct a maximal clause $l_1 \vee \sim l_2 \vee \sim l_3$using all the variables $l_1$, $l_2$, $l_3$ appearing in $S$.

*Step 2*. Delete all the clauses in $S$ containing $\sim l_1$, $l_2$, $l_3$ to obtain a new clause set $S_1$ that is obvious an empty set. So, clause set $S$ is satisfiable. Actually, let $I_D(l_1) = 0$, $I_D(l_2) = 1$, $I_D(l_3) = 1$, then $I_D(S) = 1$, and so $I_D$ is a satisfiability instance of $S$.

- **Improved method III to find a satisfiable instance**

In fact, the method (Methods I and II) for constructing satisfiability instances described above can be further improved, leading to the following approach: Let $S = \{C_1,\ldots, C_m\}$ be a satisfiable clause set in propositional logic, and let $V(S) = \{1,\ldots, n\}$ be the set of variables in $S$. Set the number of times for constructing maximal clauses about $V(S)$ to $K$. The specific steps are summarized as follows:

**Step 1**. If the number of times $k \leq K$ for randomly (or by other methods) generating a maximal clause about $V(S)$, then randomly (or by other methods) generate a maximal clause regarding $V(S)$ as $D(k)= p(1) \vee\ldots\vee p(i) \vee\ldots\vee p(n)$, where $p(i)\in \{i, \sim i\}$, $i=1,\cdots, n$, and proceed to Step 2; if the number of times $k > K$ for randomly (or by given methods) generating a maximal clause about $V(S)$, then stop.

**Step 2.** Delete the clauses containing $\sim p(1)$ from $S$ to obtain the clause set $S_1$. If $S_1=\emptyset$, then $I_D$ is a satisfiability instance of $S$; if $S_1 \neq \emptyset$: when $n > 1$, proceed to Step 3; when $n = 1$, proceed to Step 4.

**Step 3**. Repeat Step 2 for $\sim p(2)$ and $S_1$.

**Step 4**. Without affecting the clauses that have already been deleted, adjust the parts of $D(k)$ that are not involved to obtain $D^*(k)$, so as to delete the clauses in $S_1$. If $S_1=\emptyset$, then $I_{D^*(k)}$ is a satisfiability instance of $S$; otherwise, return to Step 1.

The following theorem can be obtained as a further consequence.

**Theorem 4.5** Let $C_i$ and $C_i^*$ be clauses in propositional logic, where $C_i^*$ is a non-empty subclause of $C_i$, for $i$ = 1, …, $k$. If $\bigwedge_{i=1}^{k} C_i$ is a standard contradiction, then $\bigwedge_{i=1}^{k} C_i^*$ is also a standard contradiction. In other words, any Cartesian subset of a standard contradiction is still a standard contradiction.

*Proof.* It needs only to notice that if there is complementary pair of literals among any $(x_1, \cdots, x_k) \in \prod_{i=1}^{k} C_i$, then there is also complementary pair of literals among any $(y_1, \cdots, y_k) \in \prod_{i=1}^{k} C_i^*$. ■

## 5. Triangular Standard Contradiction

As discussed in Section 3, the method for determining the properties of a clause set using a maximal contradiction proceeds as follows: first, a maximal contradiction is constructed, and then this contradiction is applied to determine whether the clause set is satisfiable or unsatisfiable. The limitation of this approach lies in its rigidity—once constructed, the composition of the maximal contradiction cannot be adjusted during the determination process. This constraint reduces the flexibility and, to some extent, the efficiency of contradiction-based deduction. To overcome this deficiency, it is necessary to develop a deductive reasoning method in which the composition of contradictions can be modified dynamically as needed during the determination process. The core theory underlying such a method is the concept of the *triangular standard contradiction* detailed in this section.

Unlike the maximal contradiction approach, where the contradiction is constructed in advance and then used to "filter" the original clause set, deduction with triangular standard contradictions integrates construction and inference simultaneously. In the maximal contradiction method, the original clause set generates the maximal contradiction, with all maximal clauses covering the original set forming the maximal clause set. The disjunction of the residual literals in each original clause (i.e., those retained after filtering) then constitutes the outcome of deduction. In contrast, triangular standard contradictions allow contradictions to be progressively constructed and adapted in parallel with the deduction process itself, offering a more flexible and efficient framework for reasoning.

### 5.1 Definition of triangular standard contradiction

An intuitive explanation of triangular standard contradictions is provided through a diagram as follows. The clauses contained within the gray triangular area in Figure 5.1 form a triangular standard contradiction. In fact, any region arbitrarily selected within the gray area of Figure 5.1, as long as it involves all clauses $D_1$, $D_2$,…, $D_k$ that constitute the triangular standard contradiction, is a standard contradiction, as established by Theorem 3.1. For example, the irregular region indicated by the curve in Figure 5.2 is one such case.

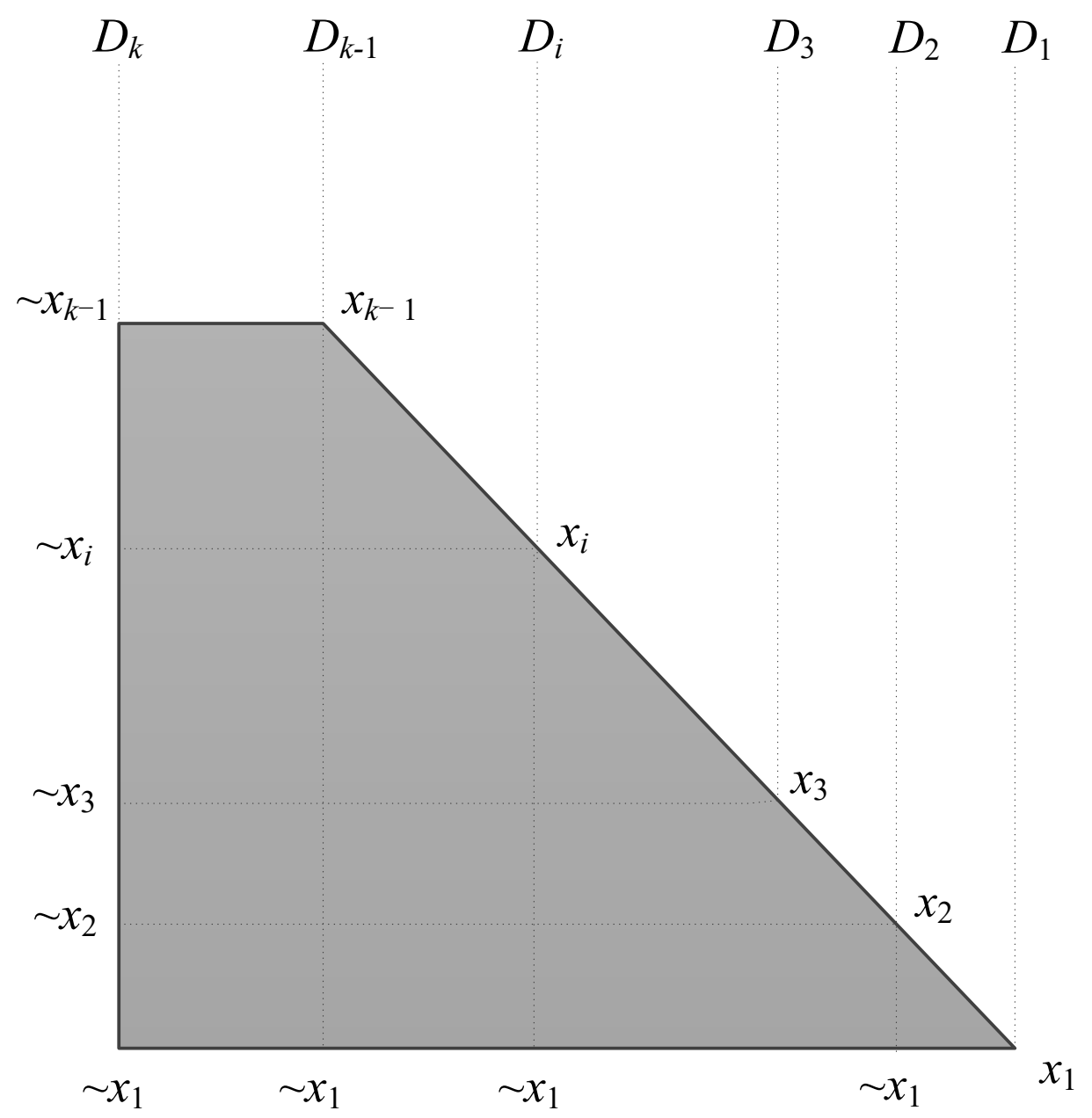

Figure 5.1 Illustration figure of triangular standard contradiction

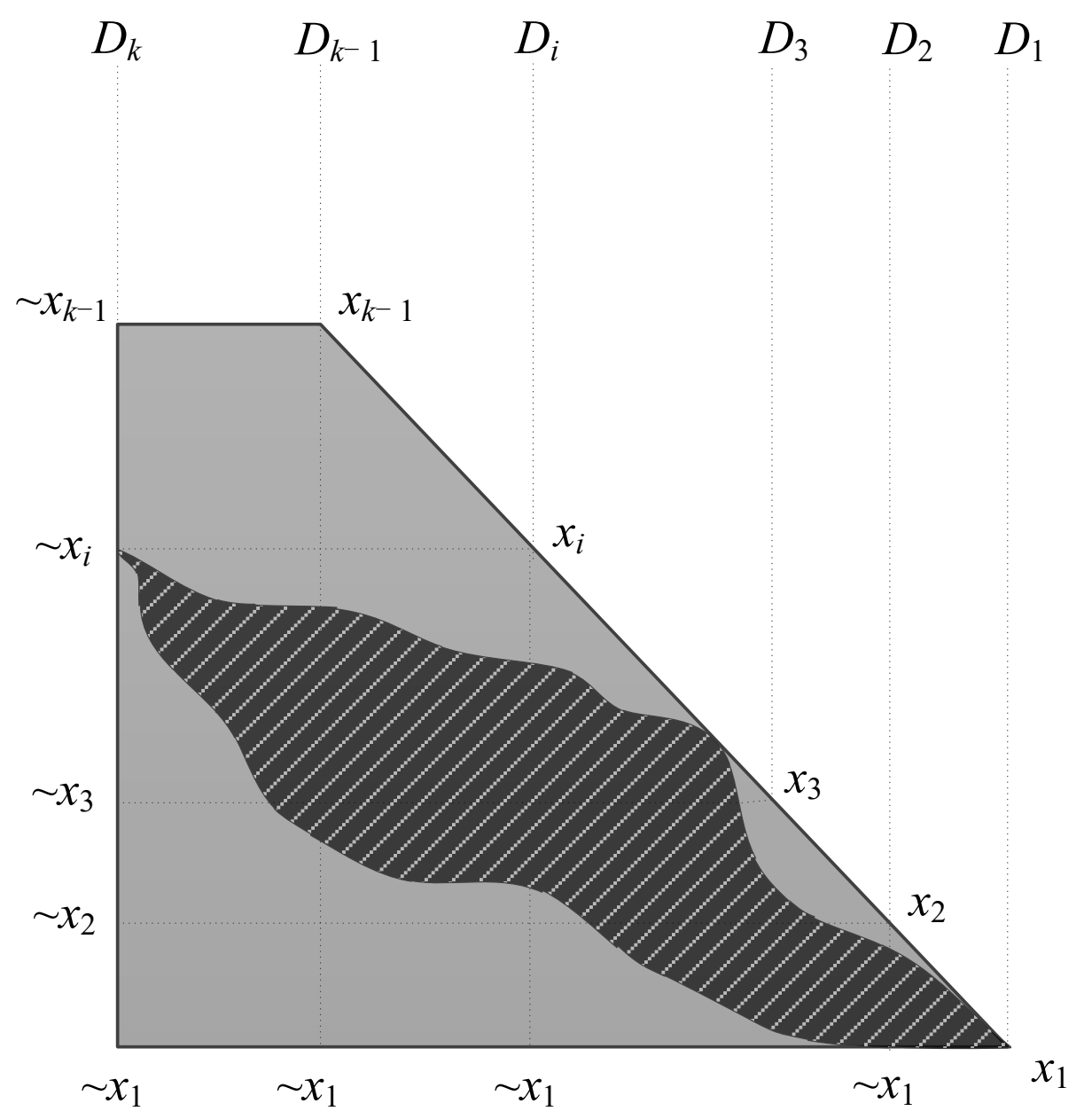

Figure 5.2 A standard contradiction

The concrete definitions and construction methods of triangular standard contradictions are provided in the following.

**Theorem 5.1** Let $x_1,\ldots, x_{k\text{-}1}$ be some literals in propositional logic or first-order logic, and clauses $D_1= x_1$, $D_2 = x_2 \vee \sim x_1,\ldots, D_t= x_t \vee \sim x_1 \vee \sim x_2 \vee\ldots\vee \sim x_{t-1},\ldots, D_{k-1} = x_{k-1} \vee \sim x_1 \vee \sim x_2 \vee\ldots\vee \sim x_{k-2}$, $D_k = \sim x_1 \vee \sim x_2 \vee\ldots\vee \sim x_{k-1}$, then $\wedge_{t=1}^{k} D_t$ is a standard contradiction. $\wedge_{t=1}^{k} D_t$ is then called a full triangular standard contradiction.

*Proof.* When $k$=2, $\wedge_{t=1}^{2} D_t = x_1 \wedge \sim x_1$, then the conclusion holds.

Suppose that the conclusion holds when $k$=$r$, i.e., $\wedge_{t=1}^{r} D_t = \wedge_{t=1}^{r-1} D_t \wedge (\sim x_1 \vee \sim x_2 \vee \cdots \vee \sim x_{r-1})$ is a standard contradiction.

We prove in the following that the conclusion holds for $k$=$r$+1.

As $D_{r+1} = \sim x_1 \vee \sim x_2 \vee \ldots \vee \sim x_r$,

$D_r = x_r \vee \sim x_1 \vee \sim x_2 \vee \ldots \vee \sim x_{r-1}$,

$D_t = x_t \vee \sim x_1 \vee \sim x_2 \vee \ldots \vee \sim x_{t-1}$, $t = r-1, \ldots, 2$,

$D_1 = x_1$.

$\wedge_{t=1}^{r+1} D_t = \wedge_{t=1}^{r} D_t \wedge D_{r+1} = \wedge_{t=1}^{r-1} D_t \wedge D_r \wedge (\sim x_1 \vee \sim x_2 \vee \ldots \vee \sim x_r)$

$= \wedge_{t=1}^{r-1} D_t \wedge (x_r \vee \sim x_1 \vee \sim x_2 \vee \ldots \vee \sim x_{r-1}) \wedge (\sim x_1 \vee \sim x_2 \vee \ldots \vee \sim x_r)$

$= (\wedge_{t=1}^{r-1} D_t \wedge x_r \wedge (\sim x_1 \vee \sim x_2 \vee \ldots \vee \sim x_r)) \vee (\wedge_{t=1}^{r-1} D_t \wedge (\sim x_1 \vee \sim x_2 \vee \ldots \vee \sim x_{r-1}) \wedge (\sim x_1 \vee \sim x_2 \vee \ldots \vee \sim x_r))$

= ①∨②

where, ① = $\wedge_{t=1}^{r-1} D_t \wedge x_r \wedge (\sim x_1 \vee \sim x_2 \vee \cdots \vee \sim x_r)$;

② = $\wedge_{t=1}^{r-1} D_t \wedge (\sim x_1 \vee \sim x_2 \vee \cdots \vee \sim x_{r-1}) \wedge (\sim x_1 \vee \sim x_2 \vee \cdots \vee \sim x_r)$.

Base on the induction assumption, $\wedge_{t=1}^{r-1} D_t \wedge (\sim x_1 \vee \sim x_2 \vee \cdots \vee \sim x_{r-1})$ is a standard contradiction, and therefore ② is a standard contradiction.

Since ①=$\wedge_{t=1}^{r-1} D_t \wedge x_r \wedge (\sim x_1 \vee \sim x_2 \vee \cdots \vee \sim x_r)$=$\wedge_{t=1}^{r-1} D_t \wedge x_r \wedge ((\sim x_1 \vee \sim x_2 \vee \ldots \vee \sim x_{r-1}) \vee \sim x_r)$

$= (\wedge_{t=1}^{r-1} D_t \wedge x_r \wedge ((\sim x_1 \vee \sim x_2 \vee \ldots \vee \sim x_{r-1})) \vee (\wedge_{t=1}^{r-1} D_t \wedge x_r \wedge \sim x_r)$

=③∨④

Base on the induction assumption, ③ is a standard contradiction and ④ is obviously a standard contradiction, and therefore ① is a standard contradiction. The conclusion holds. ■

In fact, $\wedge_{t=1}^{k} D_t$ constitutes the gray triangle area in Figure 5.1.

**Remark 5.1** In triangular contradictions, according to the proof of Theorem 5.1, it is generally not required that the set of literals on the main boundary $\{x_1, x_2, \ldots, x_{k-1}\}$ contains no complementary pairs. However, based on the construction process of triangular contradictions, it naturally follows that the set of literals on the main boundary $\{x_1, x_2, \ldots, x_{k-1}\}$ contains no complementary pairs. Special cases are handled as follows: If, when selecting a main boundary literal (e.g., $y$) for a clause $C_{i_0}$, and $y = \overline{x_{j_0}}$ where $j_0 \in \{1, \cdots, j_0 - 1\}$, then if pulling $y$ down to the expanded triangle would leave no literals above the main boundary of $C_{i_0}$, $y$ is moved down to the expanded triangle. Otherwise, $y$ is placed on the main boundary of $C_{i_0}$ and remains there until subsequent clauses are selected, at which point $x_{j_0}$ in those clauses will be moved down to the expanded triangle.

**Remark 5.2** If $y$ is placed at the main boundary line of $C_{i_0}$ as in Remark 5.1, then when the expanded triangle is completely filled, the part of the clause $C$, which comes after $C_{i_0}$ that lies within the extended triangle will contain a complementary pair, as shown in Figure 5.3.

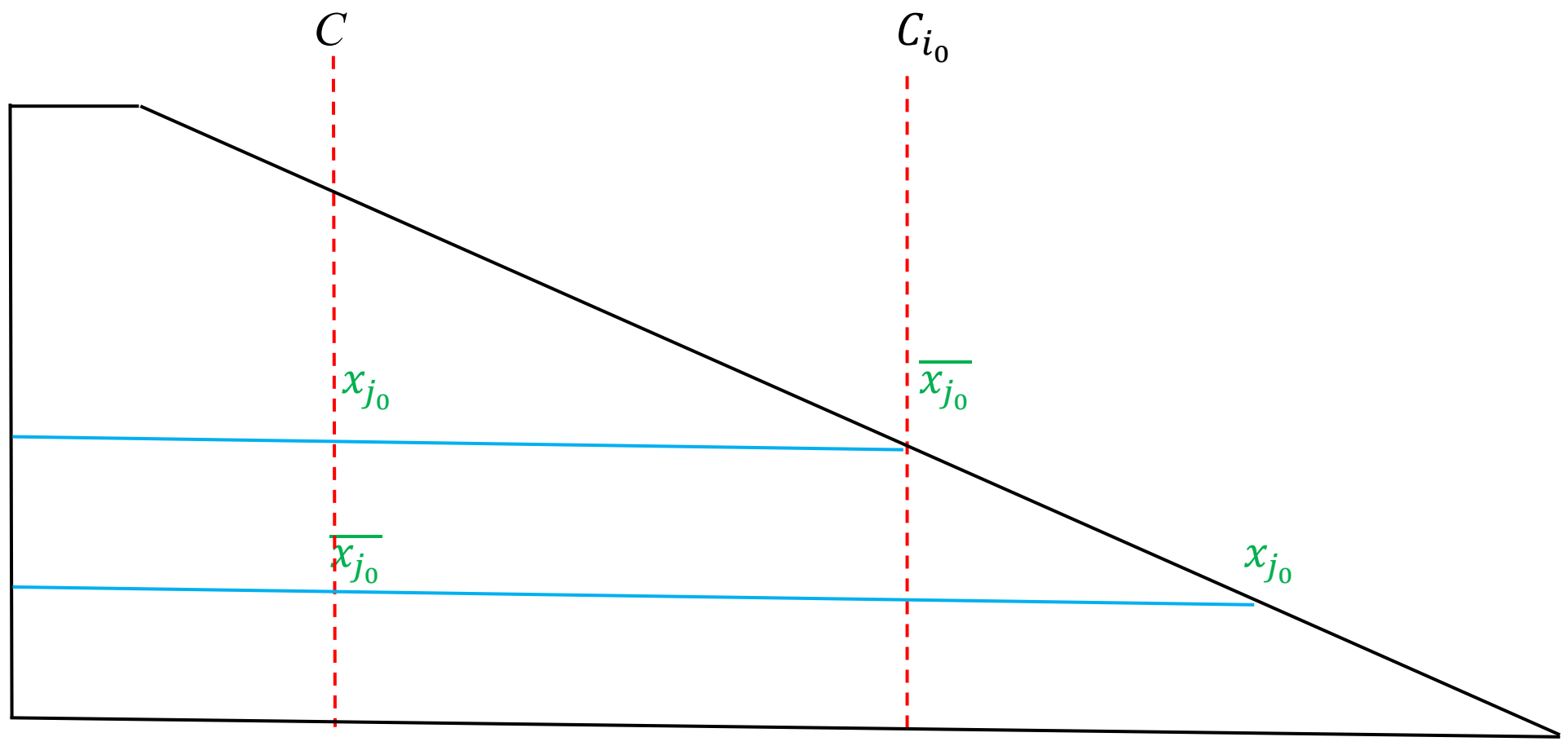

Figure 5.3. Expanded triangle containing a complementary pair on the main boundary line

**Remark 5.3** When performing expanded triangular deduction in propositional logic, since no clause in propositional logic contains complementary pairs, the situation described in Remark 5.2 does not occur. However, when performing extended triangular deduction in first-order logic, substitutions may introduce complementary pairs into clauses, thus leading to the situation described in Remark 5.2.

## 5.2 Sub-contradiction of triangular standard contradiction

This section presents some properties of sub-contradictions within triangular standard contradictions. First, an intuitive introduction to sub-contradictions is provided through Figure 5.4.

As the gray area in the triangular standard contradiction in Figure 5.4 shows, the "diagonal-line triangle", "checkered triangle", and "horizontal-line triangle" all exhibit the same structure: $\wedge_{t=1}^{T} D_t$, where $D_1 = x_1,\ldots, D_t = x_t \vee D_t^0$, and $D_t^0$ is a clause composed of some literals from $\{\sim x_1, \sim x_2,\ldots, \sim x_{t-1}\}$. $D_T = \sim x_{T-1} \vee D_T^0$, and $D_T^0$ is a clause composed of some literals from $\{\sim x_1, \sim x_2,\ldots, \sim x_{T-2}\}$. All of these marked triangles are standard contradictions.

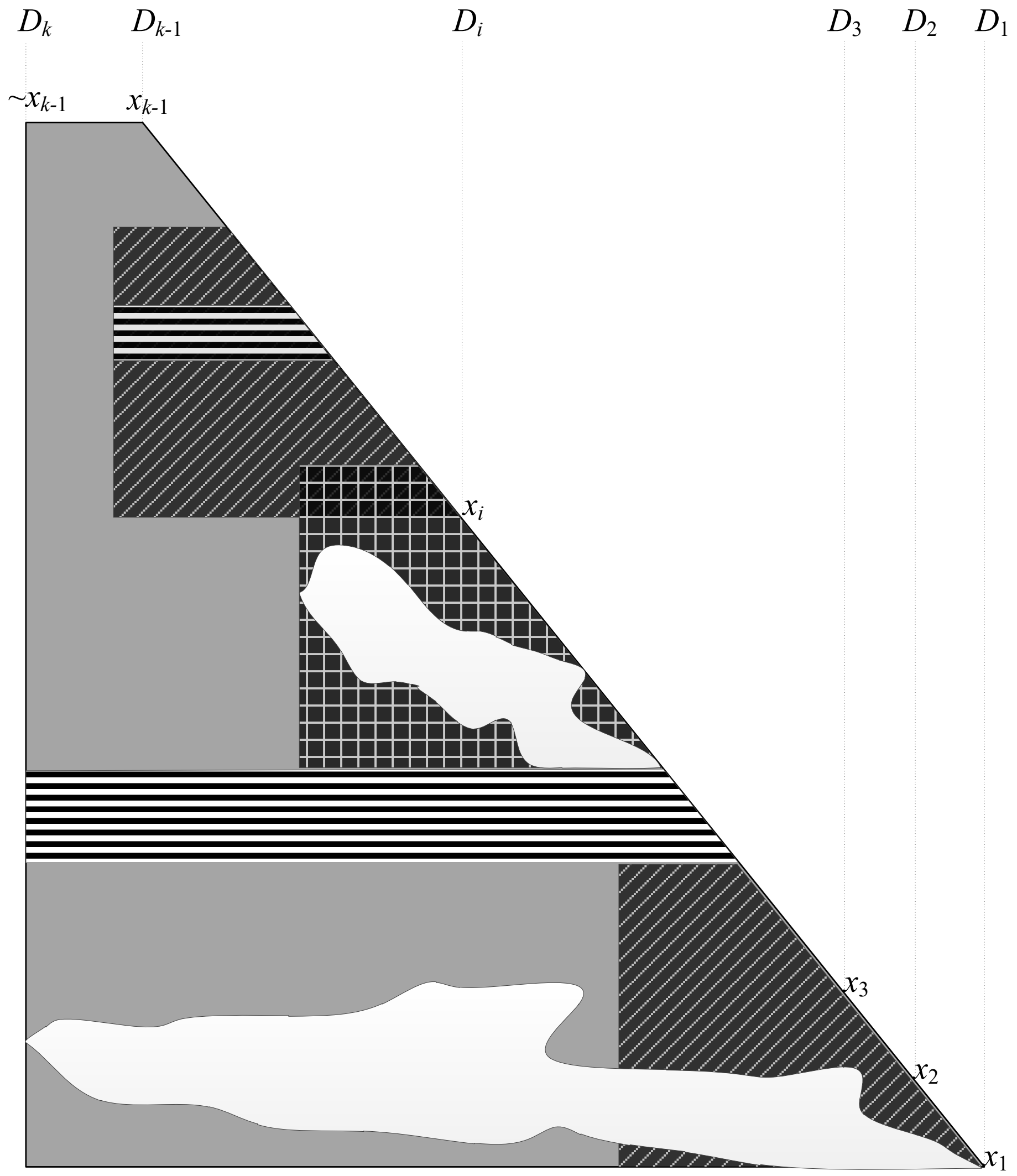

Figure 5.4 Sub-contradiction in triangular standard contradiction

**Theorem 5.2** Let $E$ be a triangle corresponding to triangular standard contradiction. For any sub-structure with the same shape of $E$, say $\wedge_{t=1}^{T} D_t$, where $D_1 = x_1, \ldots, D_t = x_t \vee D_t^0$ ($t$ = 2,…, $T$−1), and $D_t^0$ is a clause composed of some literals from $\{\sim x_1, \sim x_2, \ldots, \sim x_{t-1}\}$. $D_T = \sim x_{T-1} \vee D_T^0$, and $D_T^0$ is a clause composed of some literals from $\{\sim x_1, \sim x_2, \ldots, \sim x_{T-2}\}$, then the following conclusions hold.

(1) $\wedge_{t=1}^{T} D_t$ is a standard contradiction.

(2) If for $t$ = 1, 2,…, $T$, $D_t^*$ is a clause composed of some literals from $D_t$, then $\wedge_{t=1}^{T} D_t^*$ is a standard contradiction.

(3) Delete $D_T$ from $\wedge_{t=1}^{T} D_t$, and literal $x_{T-1}$ from $D_{T-1}$. Check whether $D_{T-1}^0, D_{T-2}^0, \cdots, D_2^0$ is empty one by one. If $D_i^0$ is the first non-empty clause, suppose that the literal with the biggest subscript in $D_i^0$ is $\sim x_{j_0}$, then delete $D_{T-1}$, $D_{T-2}, \ldots$, $D_{i+1}, x_i, D_{i-1}, D_{i-2}, \cdots, D_{j_0+1}$ consequently, and then the remaining part is a standard contradiction.

(4) For any $t_0 \in \{1,..., T-2\}$, delete clause $D_{t_0}$ and literals $x_{t_0}$, $\sim x_{t_0}$ from $\wedge_{t=1}^{T} D_t$, then the remaining part is a standard contradiction.

*Proof.* (1) If for $t = 2,..., T-1$, $D_t^0 = \sim x_1 \vee \sim x_2 \vee \ldots \vee \sim x_{t-1}$, $D_T^0 = \sim x_1 \vee \sim x_2 \vee \ldots \vee \sim x_{T-2}$, then according to Theorem 4.1, $\wedge_{t=1}^{T} D_t$ is a triangular standard contradiction. Clearly, for any $y_t \in D_t$, where $t = 1,..., T$, there exists a complementary pair among $y_1, y_2, \ldots, y_T$. Therefore, when $D_1 = x_1, \ldots, D_t = x_t \vee D_t^0$ ($t = 2, \ldots, T-1$), where $D_T^0$ is a clause composed of some literals from $\{\sim x_1, \sim x_2, \ldots, \sim x_{t-1}\}$, and $D_T = \sim x_{T-1} \vee D_T^0$, where $D_T^0$ is a clause composed of some literals from $\{\sim x_1, \sim x_2, \ldots, \sim x_{T-2}\}$, then for any $y_t \in D_t$, $t = 1,..., T$, there exists a complementary pair among $y_1, y_2, \ldots, y_T$. That is, $\wedge_{t=1}^{T} D_t$ is a standard contradiction, and the conclusion holds.

(2) Since $\wedge_{t=1}^{T} D_t$ is a standard contradiction, for any $y_t \in D_t$, $t = 1,..., T$, there exists a complementary pair among $y_1, y_2, \ldots, y_T$. Furthermore, for $t = 1,..., T$, $D_t^*$ is a clause composed of some literals in $D_t$; obviously, for any $y_t^* \in D_t^*$, $t = 1,..., T$, there exists a complementary pair in $y_1^*, y_2^*, \cdots, y_T^*$, which means $\wedge_{t=1}^{T} D_t^*$ is a standard contradiction.

(3) If $D_T$ is deleted, $D_i^0$ is the first non-empty clause of $D_{T-1}^0, D_{T-2}^0, \cdots, D_2^0$, and the literal with the biggest subscript in $D_i^0$ is $\sim x_{j_0}$. The remaining part is $\wedge_{t=1}^{j_0} D_t \wedge D_i^0$ after deleting $D_{T-1}, D_{T-2}, \cdots, D_{i+1}, x_i, D_{i-1}, D_{i-2}, \cdots, D_{j_0+1}$ consequently, where $D_1 = x_1, \ldots, D_t = x_t \vee D_t^0$ ($t = 2, \ldots, j_0$), $D_t^0$ is a clause generated by some of the literals from $\{\sim x_1, \sim x_2, \ldots, \sim x_{j_0-1}\}$. Therefore, the remaining part has the same structure with $\wedge_{t=1}^{T} D_t$, and so the conclusion holds.

(4) For $\wedge_{t=1}^{T} D_t$, where $D_1 = x_1, \ldots, D_t = x_t \vee D_t^0$ ($t = 2, \ldots, T-1$), $D_t^0$ is a clause generated by some of the literals from $\{\sim x_1, \sim x_2, \ldots, \sim x_{t-1}\}$. $D_T = \sim x_{T-1} \vee D_T^0$, $D_T^0$ is a clause generated by some of the literals from $\{\sim x_1, \sim x_2, \ldots, \sim x_{T-2}\}$. For any $t_0 \in \{1,..., T-2\}$, the remaining part is $\wedge_{t=1, t \neq t_0}^{T} D_t$ after deleting clause $D_{t_0}$ and literals $x_{t_0}$, $\sim x_{t_0}$, where $D_1 = x_1, \ldots, D_t = x_t \vee D_t^0$ ($t = 2, \ldots, T-1$, $t \neq t_0$), $D_t^0$ is a clause generated by some of the literals from $\{\sim x_1, \ldots, \sim x_{t_0-1}, \sim x_{t_0+1}, \ldots, \sim x_{t-1}\}$. $D_T = \sim x_{T-1} \vee D_T^0$, $D_T^0$ is a clause generated by some of the literals from $\{\sim x_1, \ldots, \sim x_{t_0-1}, \sim x_{t_0+1}, \ldots, \sim x_{T-2}\}$. It is obvious that $\wedge_{t=1, t \neq t_0}^{T} D_t$ has the same structure with $\wedge_{t=1}^{T} D_t$, and therefore $\wedge_{t=1, t \neq t_0}^{T} D_t$ is a standard contradiction. ■

- **Structural Analysis of Sub-Contradictions**

As shown in Figure 5.4, within the gray triangular standard contradiction, in addition to the aforementioned "diagonal-line triangle," "checkered triangle," and "dotted triangle," there are two irregular white regions. Since the checkered triangle region is also a standard contradiction, each irregular white region contains literals from every clause of a standard contradiction. Therefore, according to Theorem 3.1, the clause sets corresponding to these two irregular white regions each form a standard contradiction. Furthermore, by Definition 3.2, the clause sets corresponding to these irregular white regions are sub-standard contradictions of the respective standard contradictions. Thus, based on the shapes of the graphical representations of sub-standard contradictions within triangular standard contradictions, sub-standard contradictions can be classified into homotypic sub-standard contradictions and non-homotypic sub-standard contradictions. In fact, triangular standard contradictions exhibit the following four types of homotypic sub-standard contradictions.

**Type I: Transversely-cut type**

The upper part of the triangular standard contradiction is horizontally cut along the row where $x_i$ is located, that is, the standard contradiction formed by the grid part in Figure 5.5, which may be denoted as $H_1$ for convenience.

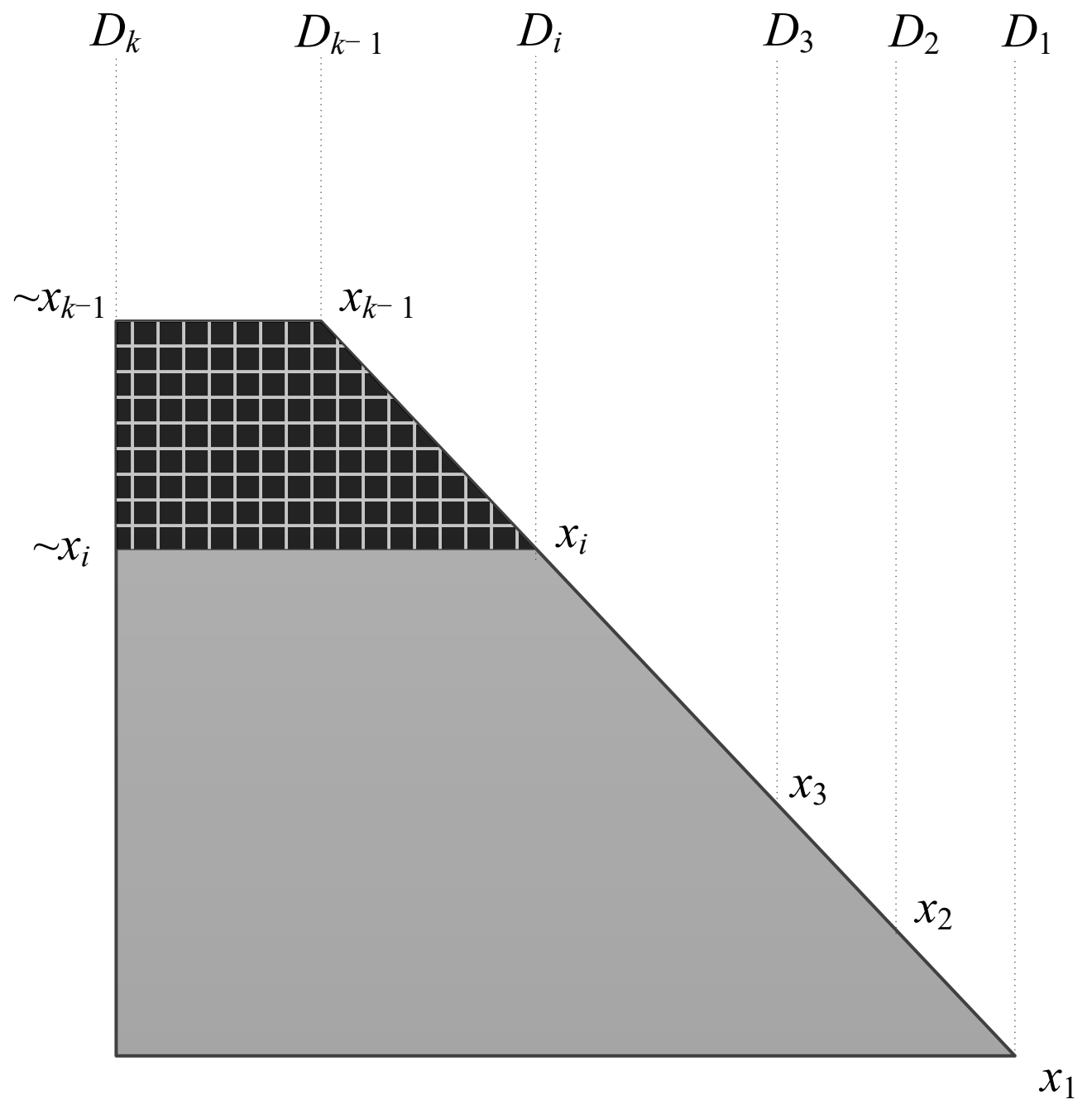

Figure 5.5 Transversely-cut triangular standard sub-contradiction

That is, $H_1 = \wedge_{t=i}^{k} D_t^*$, where $D_i^* = \{x_i\}$, $D_t^* = \{x_t\} \cup \{\sim x_j | j = i, \cdots, t-1\}$, $t = i+1, \ldots, k-1$, $D_k^* = \{\sim x_j | j = i, \cdots, k-1\}$, $x_j$ is the literal of clause set $S$.

**Type II: Vertically-cut type**

Cut out the right part of the triangular standard contradiction vertically along the column where $D_i$ is located, i.e., the standard contradiction $H_2$ formed by the grid part in Figure 5.6.

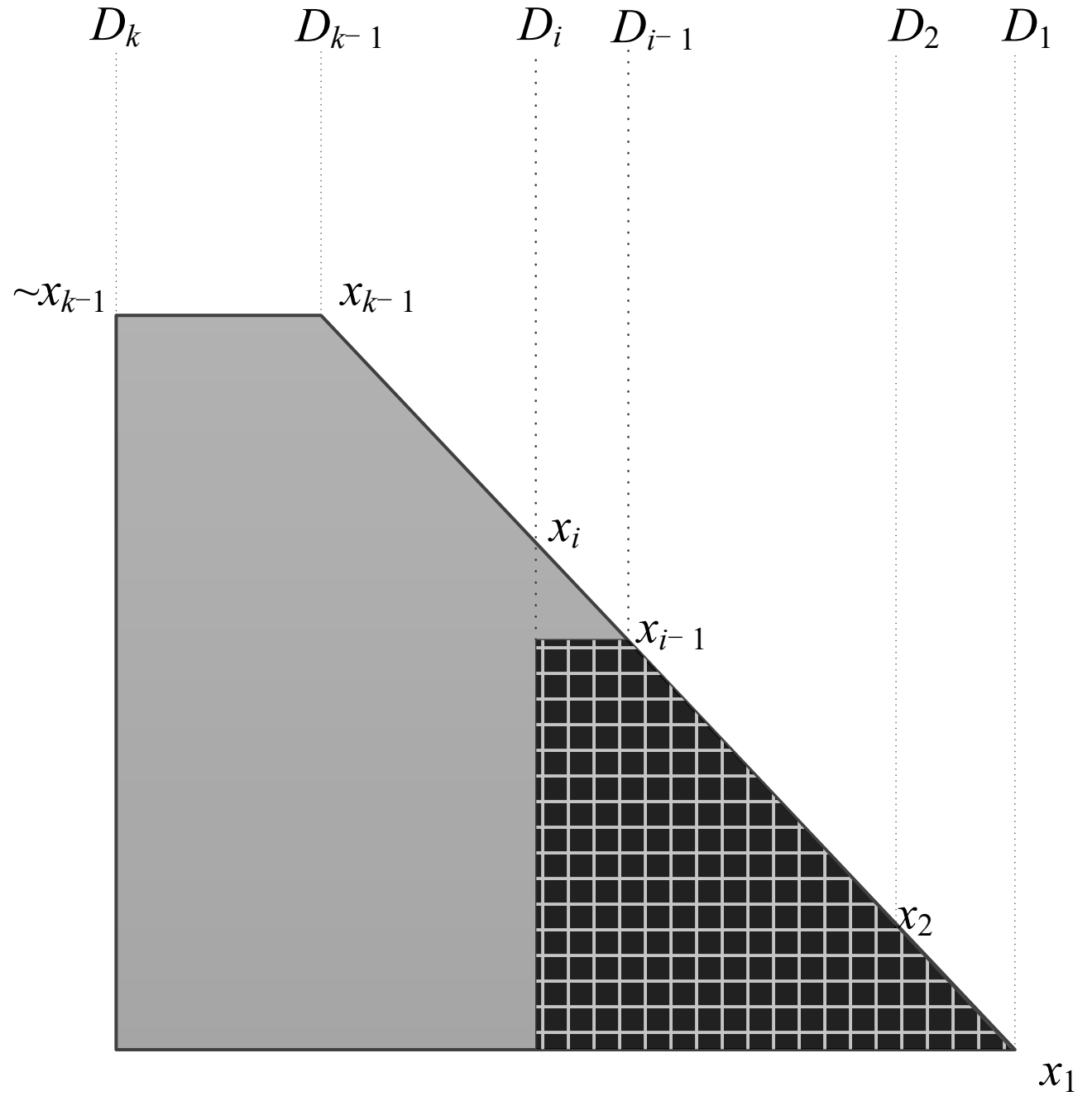

Figure 5.6 Vertically-cut triangular standard sub-contradiction

That is, $H_2 = \wedge_{t=1}^{i} D_t^*$, where $D_1^* = D_1 = \{x_1\}$, $D_t^* = D_t = \{x_t\} \cup \{\sim x_j | j = 1, \cdots, t-1\}$, $t = 2, \ldots, i-1$, $D_k^* = \{\sim x_j | j = 1, \cdots, i-1\}$, $x_j$ is the literal of clause set $S$.

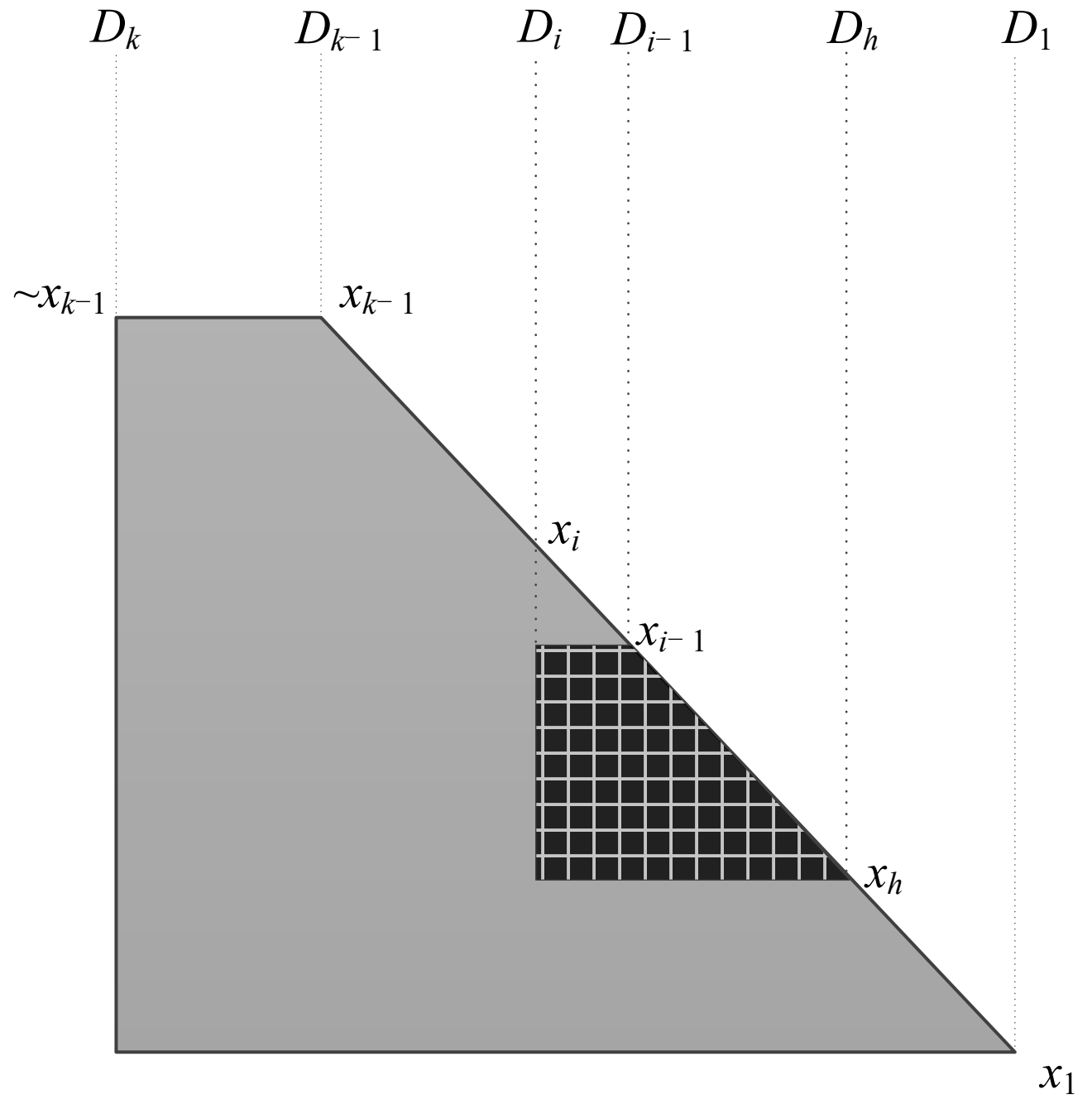

Figure 5.7 Middle type triangular standard sub-contradiction

**Type III: Middle type**

Along the row where $x_h$ is located for a horizontal cut and the column where $D_i$ is located for a vertical cut, the contradiction enclosed together with the diagonal line, that is, the standard contradiction $H_3$ formed by the grid part in Figure 5.7.

That is, $H_3 = \wedge_{t=h}^{i} D_t^*$, where $D_h^* = \{x_h\}$, $D_t^* = \{x_t\} \cup \{\sim x_j | j = h, \cdots, t-1\}$= , $t$=2,…, $i$–1, $D_i^*$={~$x_j$ | $j$ = $h$,..., $i$–1}, 1 ≤ $h$ ≤ $i$–1, $x_j$ is the literal of clause set $S$.

**Type IV: Deletion type**

Delete the columns where $D_h$ and $D_i$ are located and the rows where $x_h$ and $x_i$ are located. The remaining portion forms the standard contradiction, specifically the standard contradiction $H_4$ obtained after removing the white parts in Figure 5.8.

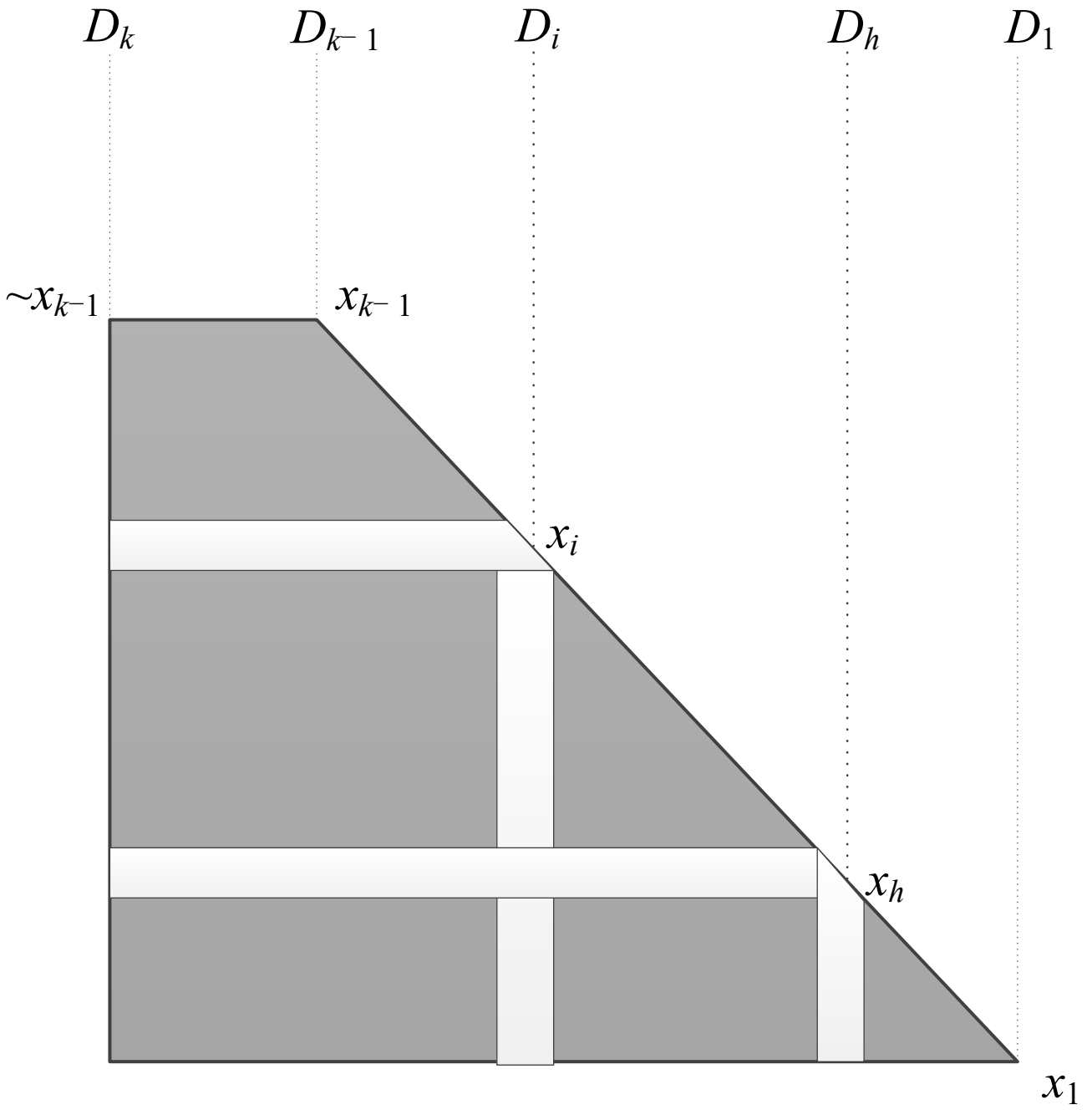

Figure 5.8 Deletion type triangular standard sub-contradiction

That is, $H_4 = \wedge_{t=1, t\neq h,i}^{k} D_t^*$, where $D_1^* = D_1 = \{x_1\}$, $D_t^* = \{x_t\} \cup \{\sim x_j | j = 1, \cdots, t-1, j \neq h, i\}$= , $t$ = 2,…, $k$–1, $t \neq h$, $i$, $D_k^* = \{\sim x_j \,|j = 1, \ldots, k\text{–}1, j \neq h, i\}$, $x_j$ is the literal of clause set $S$.

- **Invariant under scaling about standard contradiction**

The standard contradiction is invariant under scaling as detailed below:

1) For a clause set $S$={$C_1$,…,$C_m$}, if the constructed standard contradiction is $\wedge_{i=1}^{m} C_i$, then it can be expanded to a bigger standard contradiction $\wedge_{i=1}^{m} (C_i \vee l) \wedge \wedge_{i=1}^{m}(C_i \vee \sim l)$, where $l$ is any literal in the clause set.

In fact, for any $(x_1,\ldots, x_m, x_{m+1},\ldots, x_{2m}) \in \prod_{i=}^{m}(C_i \vee l) \times \prod_{i=}^{m}(C_i \vee \sim l)$:

(i) If $(x_1,\ldots, x_m) \in \prod_{i=}^{m} C_i$ or $(x_{m+1},\ldots, x_{2m}) \in \prod_{i=}^{m} C_i$, then there are complementary pairs of literals among $(x_1,\ldots, x_m)$ and $(x_{m+1},\ldots, x_{2m})$ because $\wedge_{i=1}^{m} C_i$ is a standard contradiction. Therefore, there is complementary pair of literals among $(x_1,\ldots, x_m, x_{m+1},\ldots, x_{2m})$.

(ii) If $(x_1,\ldots, x_m) \notin \prod_{i=}^{m} C_i$ and $(x_{m+1},\ldots, x_{2m}) \notin \prod_{i=}^{m} C_i$, then there exists $x_{i1} \in \{x_1,\ldots,x_m\}$ and $x_{i2} \in \{x_{m+1},\ldots, x_{2m}\}$, $x_{i1}= l$ and $x_{i2}=\sim l$, furthermore, there is complementary pair of literals among $(x_1,\ldots, x_m, x_{m+1},\ldots, x_{2m})$, say, $l$ and $\sim l$. Therefore, there is complementary pair of literals among $(x_1,\ldots, x_m, x_{m+1},\ldots, x_{2m})$.

2) For clause set $S=\{C_1,\ldots,C_m\}$, if the constructed standard contradiction is $\wedge_{i=1}^{m} C_i$, then for any clause set $S_2=\{D_1,\ldots, D_n\}$, it can be expanded to a bigger standard contradiction $\wedge_{i=1}^{m} C_i \wedge \wedge_{j=1}^{n} D_j$.

In fact, for any $(x_1,\ldots, x_m, x_{m+1},\ldots, x_{2m}) \in \prod_{i=}^{m} C_i \times \prod_{j=1}^{n} D_j$, there are complementary pairs of literals among $(x_1,\ldots, x_m)$ because $\wedge_{i=1}^{m} C_i$ is a standard contradiction. Therefore, there are complementary pairs of literals among $(x_1,\ldots,x_m, x_{m+1},\ldots,x_{2m})$.

3) For clause set $S=\{C_1,\ldots,C_m\}$, if the constructed standard contradiction is $\wedge_{i=1}^{m} C_i$, then is can shrink to a smaller standard contradiction $\wedge_{i=1}^{m} [(C_i - l)\text{or}(C_i - \sim l)] = \wedge_{i=1}^{m} (C_i - \{l, \sim l\})$, where $l$ and $\sim l$ are any literals in the clause set.

Actually, it is noted that, for each clause $C_i$, there is only one literal from $l$ and $\sim l$ can be contained in $C_i$. Therefore, for any $(x_1,\ldots,x_m) \in \prod_{i=1}^{m}[(C_i - l)\text{or}(C_i - \sim l)] = \prod_{i=1}^{m}[(C_i - \{l, \sim l\}]$, i.e., $(x_1,\ldots,x_m) \in \prod_{i=1}^{m} C_i$, but $(x_1,\ldots,x_m) \notin \prod_{i=1}^{m}\{l, \sim l\}$.

Because $\wedge_{i=1}^{m} C_i$ is standard contradiction, there is no complementary pair of literals among $(x_1,\ldots, x_m)$.For triangular standard contradiction, it is actually Type IV shown above.

Based on the definition of standard contradiction, we have the following theorem.

**Theorem 5.3** Let $S$ be a clause set, $E = \wedge_{t=1}^{k} D_t$ is a triangular standard contradiction, where $D_1=\{x_1\}$, $D_t= \{x_t\} \cup \{\sim x_j \mid j =1,\ldots, t-1\}$, $t = 2,\ldots, k-1$, $D_k =\{\sim x_j \mid j =1,\ldots, k-1\}$, and $x_j$ is the literal of $S$ for $j =1,\ldots, k-1$. Then $\wedge_{D \in S-\{D_1,\ldots,D_k\}} D \wedge \wedge_{t=1}^{k} D_t$ is the standard contradiction with the maximum number of clauses in $S$ containing $E$.

In a triangular standard contradiction, after deleting some literals, the remaining part may still form a standard contradiction. Therefore, to study the number of substandard contradictions contained within a triangular standard contradiction, with the requirement that the standard sub-contradiction contains the same number of clauses as the triangular standard contradiction, the following example of a triangular standard contradiction composed of $n$ clauses $D_1, D_2,\ldots, D_n$ is used to illustrate this problem.

In fact, the triangular standard contradiction formed by $D_1$, $D_2$,…, $D_n$, denoted as Δ for convenience, corresponds to the following Figure 5.9.

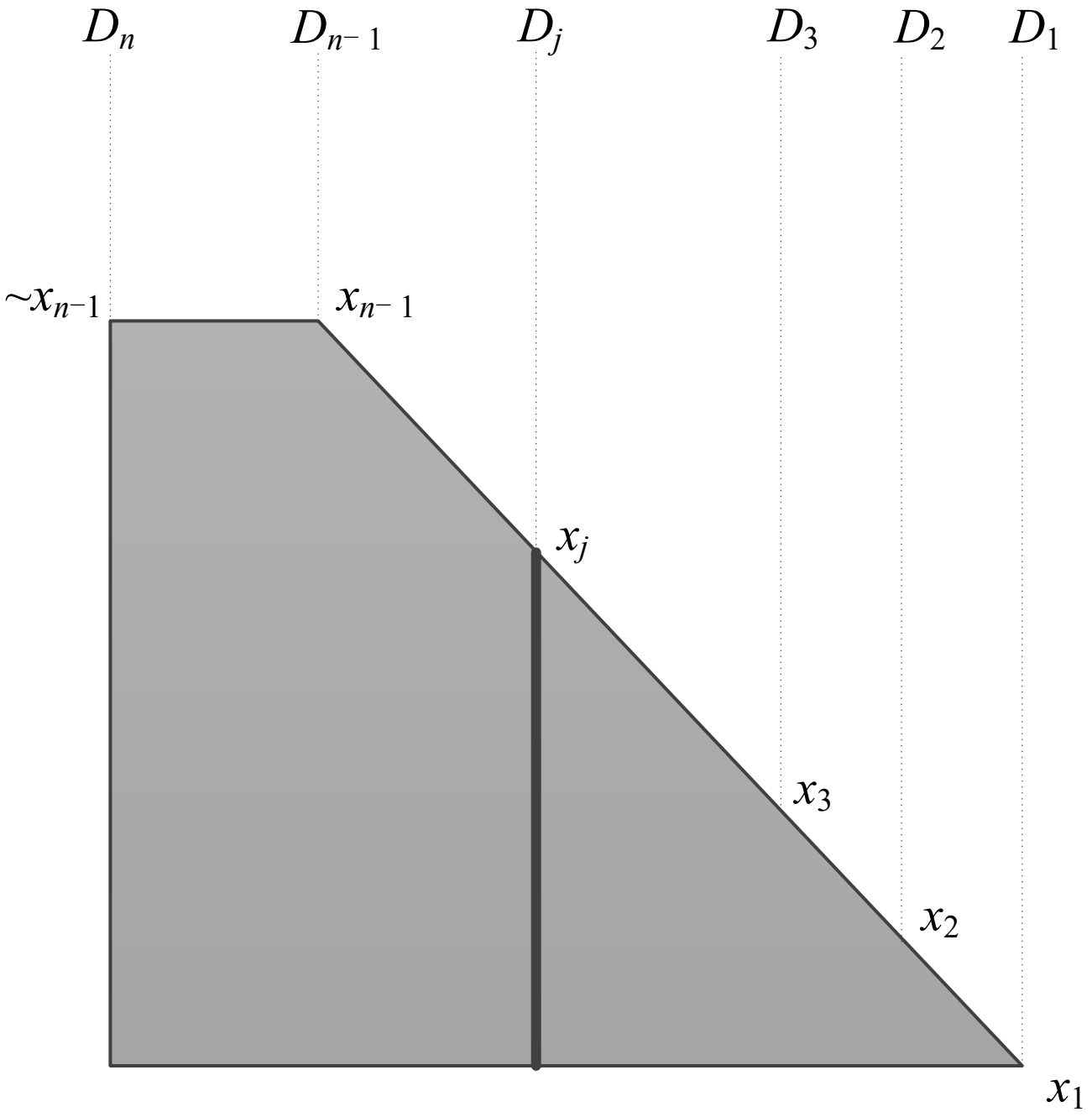

Figure 5.9 Deletion type triangular standard sub-contradiction

Since the number of clauses contained in the standard sub-contradiction is the same as that in Δ, for $j$ = 1, 2, …, $n$-1, the number of types of literals placed in $D_j$ is as follows:

$$C_j^1 + C_j^2 + \ldots + C_j^j = \sum_{k=1}^{j} C_j^k \qquad (5.1)$$

where $C_j^k$ denotes the number of ways to place literals when only $k$ out of the $j$ positions of $D_j$ have literals, where $1 \le k \le j$. Since the number of ways to place literals in these $j$ positions should be equal to the number of non-empty subsets of $\{\sim x_1, \ldots, \sim x_{j-1}, x_j\}$, this is thus a combination problem. For the last clause $D_n$, as $D_n$ corresponds to $n$-1 positions, the number of ways to place literals in $D_n$ is $\sum_{k=1}^{n-1} C_{n-1}^k$. Additionally, it should be noted that the relationship between the clauses is a "conjunction" relationship; when calculating the total number of ways to place literals across all clauses, "conjunction" is reflected as a "multiplication" relationship. Therefore, the total number of ways to place literals is as follows.

$$CN(n) = \prod_{j=1}^{n-1} \sum_{k=1}^{j} C_j^k \times \sum_{k=1}^{n-1} C_{n-1}^k \qquad (5.2)$$

That is, the triangular standard contradiction of contains $CN(n)$ distinct standard sub-contradictions.

Suppose the maximum standard contradiction contains $n$ literals, then this maximum standard contradictory body has $2^n$ clauses, with each clause containing $n$ literals. The number of standard sub-contradictions it contains, denoted as $MSC(n)$, is calculated as follows.

$$MSC(n) = \left(\sum_{i=1}^{n} C_n^i\right)^{2^n} \tag{5.3}$$

In binary resolution, only the simplest contradiction, namely a complementary pair, can be separated at a time. This demonstrates that in contradiction separation based dynamic automated reasoning, a large number of contradictions can be separated simultaneously. This sufficiently illustrates that contradiction separation based dynamic automated deduction possesses strong reasoning capabilities.

## 6. Conclusions and Future Works

This paper has advanced the theory and methodology of contradiction-separation-based dynamic multi-clause automated deduction by focusing on the systematic construction of standard contradictions. Building upon the theoretical foundations established in 2018 [8], we introduced two principal forms of standard contradictions: the maximal triangular standard contradiction and the triangular-type standard contradiction. For maximal contradictions, we presented their definition, construction method, and theoretical properties, demonstrating their effectiveness in determining both satisfiability and unsatisfiability of clause sets, and in constructing satisfiable instances when applicable. To address the rigidity of maximal contradictions, we then developed the theory of triangular standard contradictions, which support the dynamic adjustment of contradiction composition during the deduction process.

Furthermore, we analyzed the structural properties of triangular standard contradictions by identifying and classifying their sub-contradictions into homotypic and non-homotypic types. Explicit formulas were derived for computing the number of sub-contradictions embedded within both maximal and triangular standard contradictions, thereby providing tools for quantitative structural analysis. Collectively, these results establish a methodological foundation for extending automated reasoning beyond the intrinsic limitations of classical binary resolution, enabling more expressive, efficient, and robust forms of logic-based deduction.

Looking ahead, future work will concentrate on the progressive development of dynamic deduction methods grounded in contradiction separation, particularly those that integrate multiple forms of standard contradictions in a unified framework. We also aim to investigate practical implementations of these methods in state-of-the-art ATP systems and to evaluate their performance across large-scale benchmark libraries such as TPTP and CASC competitions. Ultimately, these efforts will contribute to building the next generation of trustworthy AI reasoning systems that are both theoretically sound and practically powerful.